\documentclass[twocolumn, trackchanges]{aastex701} %, linenumbers, trackchanges
\definecolor{trackchange}{cmyk}{0,1.0,0.6,0.2}

\usepackage[T1]{fontenc}
\usepackage{ae,aecompl}

\usepackage[normalem]{ulem} %Strikethrough support

\usepackage{graphics,epsf}
\usepackage[utf8]{inputenc}
\usepackage{amsmath}                % American Mathematical Society package
\usepackage{amsfonts}               % American Mathematical Society fonts
\usepackage{amssymb}                % American Mathematical Sociy symbol
\usepackage{epsfig}                 % EPS figures
\usepackage{graphicx}
\usepackage{float}
\usepackage{color}
\usepackage{multirow}               % double row table entries
\usepackage{hyperref}
\usepackage{xspace}
\usepackage{rotating}

\hypersetup{
    colorlinks=true,
    linkcolor=red,   
    urlcolor=cyan}

\usepackage[para,online,flushleft]{threeparttable}
\usepackage{lineno}

\newcommand{\mesa}{\textsc{mesa}}

\newcommand{\kms}{{~\rm km\; s^{-1}}}

\newcommand{\cm}{{~\rm cm}}
\newcommand{\km}{{~\rm km}}
\newcommand{\s}{{~\rm s}}

\newcommand{\g}{{~\rm g}}
\newcommand{\G}{{~\rm G}}
\newcommand{\K}{{~\rm K}}
\newcommand{\erg}{{~\rm erg}}

\newcommand{\Mo}{~M_\odot}

\shorttitle{Jittering jets in CCSNe}
\shortauthors{Wang, Shishkin, Soker}

\begin{document}

\title{The explosion time of the jittering jet explosion mechanism (JJEM)}
%\date{April 2025}

\author[0009-0004-9646-5271]{Nikki Yat Ning Wang}
\affiliation{Department of Physics, Technion - Israel Institute of Technology, Haifa, 3200003, Israel}
\email{nikki.wang@campus.technion.ac.il}
\correspondingauthor{Nikki Yat Ning Wang}

\author[0000-0002-9444-9460]{Dmitry Shishkin}
\affiliation{Department of Physics, Technion - Israel Institute of Technology, Haifa, 3200003, Israel}
\email{s.dmitry@campus.technion.ac.il}

\author[0000-0003-0375-8987]{Noam Soker} 
\affiliation{Department of Physics, Technion - Israel Institute of Technology, Haifa, 3200003, Israel}
\email{soker@physics.technion.ac.il}

\begin{abstract}
We compare the expected explosion time of core-collapse supernovae (CCSNe) in the jittering jets explosion mechanism (JJEM) to that in the neutrino-driven (the delayed neutrino) mechanism, and conclude that for most, but not all, stellar models, even when the neutrino-driven mechanism might have led to an explosion, the JJEM starts to operate before the neutrino mechanism fully revives the stalled shock. To achieve this goal, we conduct comprehensive one-dimensional stellar evolution simulations using the code \textsc{mesa} to determine the expected explosion time according to the JJEM, which corresponds to the time when the inner convective layer of the core is accreted onto the newly born neutron star (NS). We present results for many hydrogen-rich, hydrogen-stripped, and helium+hydrogen-stripped envelope CCSN progenitor models. We find that most, but not all, of our models in the initial mass range of $M_{\rm ZAMS}=12M_\odot$ to $M_{\rm ZAMS}=40M_\odot$, have at least one strong convective zone inside the mass coordinate that is the maximum baryonic mass of an NS, $m \simeq 2.65\ M_\odot$, which can facilitate the JJEM. This study contributes to establishing the JJEM as a viable explosion mechanism of CCSNe.    
\end{abstract} 
% \keywords{Core-collapse supernovae; Stellar jets; Massive stars}
% https://astrothesaurus.org/concept-select/

% ============================================
\section{INTRODUCTION}
\label{sec:Intro}
% ============================================
% cite: author (year) | citep: (author year) | citealt: author year

The identification of point-symmetric morphologies in about 20 core-collapse supernova (CCSN) remnants (CCSNRs) in recent years marks a breakthrough in the study of the jittering jets explosion mechanism (JJEM) of CCSNe that predicts such morphologies (\citealt{Soker2024Rev, Soker2024RevUniv} for reviews). As such, the JJEM became one of two heavily studied CCSN explosion mechanisms; the second being the neutrino-driven mechanism (the delayed neutrino explosion mechanism). Within the framework of JJEM, neutrino heating might boost the jets' energy in the JJEM \citep{Soker2022nu}, but jets supply most of the explosion energy. 

{{There are common processes to the two mechanisms (e.g., \citealt{Soker2025Learning}). Both mechanisms require the pre-collapse core to possess vigorous convection: to set perturbations in the neutrino-driven mechanism, and to seed angular momentum fluctuations in the JJEM. As well, in the JJEM, once jets propagate through the stalled shock, neutrino heating boosts their energy \citep{Soker2022_Boosting}. Namely, neutrino heating plays an important role in the JJEM, but it is not the primary driver of the explosion. }}

Because in this paper we compare one of the properties of the two competing explosion mechanisms, the explosion time, we open by detailing the approaches that researchers of the two mechanisms take to study and support the mechanisms.
Researchers of the neutrino-driven mechanism in recent years have focused their studies on the three-dimensional simulations of the explosion process by the delayed-neutrino mechanism, from core collapse to shock revival, and have attempted to show that some observations are compatible with some predictions of the neutrino-driven mechanism (e.g., \citealt{Bambaetal2025CasA, Bocciolietal2025Si, EggenbergerAndersenetal2025, Huangetal2025, Imashevaetal2025, Janka2025, Laplaceetal2025, Maltsevetal2025, Morietal2025, Mulleretal2025, Nakamuraetal2025, OrlandoJankaetal2025B, SykesMuller2025, ParadisoCoughlin2025, Schneideretal2025, Vinketal2025, WangBurrows2025, Willcoxetal2025, Mukazhanov2025, Antonietal2025, FangQetal2025, Raffeltetal2025, Vartanyanetal2025, Calvertetal2025, Giudicietal2026, LuoZhaKajino2026, Murphyetal2026, Orlando2026, Rusakovetal2026, VarmaMuller2026, Wessonetal2026}). However, the neutrino-driven mechanism has no explanation for point-symmetric CCSN morphologies (\citealt{Soker2025Learning} for a recent comparison of the two explosion models in that regard).  
The magnetorotational explosion mechanism posits that a pair of opposite jets, aligned along a fixed axis, propel the star's explosion. The fixed angular momentum axis along which the accretion disk launches the jets requires a rapidly rotating pre-collapse core (e.g., \citealt{Shibagakietal2024, ZhaMullerPowell2024, Shibataetal2025, PanLi2026} and references therein to much older papers), a very rare property (e.g., \citealt{Shibataetal2025, Mannoetal2026}). The formation of a black hole in collapsars (e.g., \citealt{BoppGottlieb2025, Gottliebetal2025}, for recent papers) is also very rare. The magnetorotational explosion mechanism can explain point symmetry only in rare cases of CCSNRs. However, point-symmetric CCSNRs are common.  The neutrino-driven mechanism attributes the extra energy of some CCSNe with explosion energies above what the neutrino mechanism can supply to an energetic magnetar, i.e., a rapidly spinning neutron star (NS). However, many magnetar models of such energetic (superluminous) CCSNe have explosion energies of $E_{\rm exp} \gtrsim 3 \times 10^{51} \erg$ (e.g., \citealt{Aguilaretal2025, Orellanaetal2025}, for some of the most recent examples) that imply explosion by jets (e.g., \citealt{SokerGilkis2017, Kumar2025}). (For the relation between the explosion mechanisms and the extra energy sources in CCSNe see \citealt{Soker2025G11}.)

Recent studies in the JJEM have focused on identifying point-symmetric morphologies, e.g., in the Crab Nebula \citep{ShishkinSoker2025Crab}, including puzzling CCSNRs, e.g., 3C 397 \citep{Klimovetal2026}. The method these studies have been using is new to the study of CCSNe. However, it is an old one that has yielded a deep understanding of the shaping of planetary nebulae by jets (e.g.,  \citealt{Morris1987, Soker1990AJ, SahaiTrauger1998, AkashiSoker2018,   EstrellaTrujilloetal2019, Tafoyaetal2019, Balicketal2020,   GarciaSeguraetal2020, GarciaSeguraetal2021, Clairmontetal2022, RechyGarciaetal2020, Danehkar2022, MoragaBaezetal2023, Ablimit2024, Derlopaetal2024, Mirandaetal2024, Sahaietal2024}), including precessing jets (e.g., \citealt{Guerreroetal1998, Mirandaetal1998, Sahaietal2005, Boffinetal2012, Sowickaetal2017, RechyGarciaetal2019, Guerreoetal2021, Clairmontetal2022}). 

The method starts with a careful eye inspection and qualitative classification based on morphology (e.g., \citealt{Balick1987, Parkeretal2006, Sahaietal2007, Kwok2024}). Although qualitative, this method enables the robust identification of jet-shaped structures (e.g., \citealt{SahaiTrauger1998}). The identification of jet-shaped structural features motivates jet-shaped numerical simulations. The qualitative and partially quantitative comparisons of simulations with observations (e.g., \citealt{GarciaSeguraetal2021, GarciaSeguraetal2022, GarciaSeguraetal2025}) established the major role of jets in shaping and powering outflows in planetary nebulae. The systematic comparison of 3D hydrodynamical simulations of CCSN jets in the framework of the JJEM is new (e.g., \citealt{Braudoetal2025, Braudoetal2026, SokerAkashi2025, AkashiSoker2026a, AkashiSoker2026bG11}). 
Comparing the similarity of some CCSNRs to planetary nebulae (e.g., \citealt{Akashietal2018}) also strengthens the case for jet powering in many CCSNRs.

CCSNe interact with the circumstellar material lost by the CCSN progenitor (e.g., \citealt{Chiotellisetal2021, ChiotellisZapartasMeyer2024, Velazquezetal2023, Meyeretal2022, MeyerDetal2024}) and with the interstellar medium (e.g., \citealt{Wuetal2019, YanLuetal2020, LuYanetal2021, MeyerMelianietal2024}). These interactions in general smear the point-symmetric morphology, although in some cases might add point-symmetric morphological features, but only in the outer SNR regions, as some studies suggest for type Ia supernova remnant (e.g., \citealt{Soker2024SNRG1903}).  
Other processes contribute to the smearing of point symmetry, including the NS natal kick, instabilities during the explosion, post-explosion processes such as a pulsar wind nebula (if it exists), and heating from the reverse shock and radioactive decay. In many cases, these smearing processes make point-symmetric morphology identification challenging, if not impossible. Magnetic fields in the ISM, as in the scenario of \cite{YuHandFangJ2025}, might explain one pair of ears in rare SNRs, but are unable to explain point-symmetrical structures, either in SNRs or in planetary nebulae; at best, they might account for rare cases.

Studies of the JJEM cannot track the evolution from core collapse to jet launching. Instead, the different evolutionary phases are studied separately. This type of study is done, for example, in examining the shaping of planetary nebulae, where the binary interaction that leads to accretion disk formation  (e.g., \citealt{DeMarcoetal2025} for a review) is separated from the study of the planetary nebula shaping processes at a later phase  (e.g., \citealt{AkashiSoker2021, GarciaSeguraetal2022}). In general, the simulations of mass transfer in binary systems are extremely complicated (e.g., \citealt{Kashietal2022}), in particular when accretion disks are formed (e.g., \citealt{Kashi2023}), and do not allow for the inclusion of jet launching. The same holds for the JJEM because the launching of jets by the intermittent accretion disks around the newly born NS in the JJEM involves violent magnetic field reconnection events. Current numerical codes cannot resolve small-scale magnetic field reconnection or simulate the JJEM \citep{Soker2025Learning}. Therefore, simulations for the powering of the explosion and the shaping of the ejecta (e.g., \citealt{AkashiSoker2022CCSN, AbdikamalovBeniamini2025, Braudoetal2025}) differ from the tools used to study the source of the angular momentum in intermittent accretion disks.   

Some JJEM studies focus on demonstrating that the pre-collapse core's convection zones exhibit large fluctuations in both velocity and angular momentum, which are characterized by significant variations in magnitude and direction. Instabilities above the NS (for instabilities and turbulence in the gain region, see, e.g., \citealt{Abdikamalovetal2016, KazeroniAbdikamalov2020, Buelletetal2023}) amplify these stochastic angular momentum fluctuations (e.g., \citealt{WangShishkinSoker2024}). These JJEM studies simulated hydrogen-rich progenitors, i.e., type II CCSNe, and demonstrated large convective velocity fluctuations in the core's silicon- and oxygen-burning zones. The exact convection structure depends on accurately following a large number of nuclear reactions at the final phases before collapse, which is a difficult task by itself (e.g., \citealt{Grichener_etal_2025_NuclearNN}).

This study aims to provide an explosion-time comparison between the two competing theoretical explosion mechanisms by examining pre-collapse convective and velocity fluctuations in progenitors that have lost most of their hydrogen and helium, specifically stripped-envelope CCSNe. We use 1D simulations (Section \ref{sec:NumScheme}), but note that 3D simulations find larger convective velocities and fluctuations than 1D calculations. 
We present the results for the stripped-envelope models in Section \ref{sec:results_models}.
Both single and binary evolution can form hydrogen and helium-depleted massive stars (e.g., \citealt{Shenaretal2020} on single star evolution and \citealt{Hiraietal2020} on binary interaction; see \citealt{Gilkisetal2025} for a recent thorough study of binary interaction, and \citealt{Kawashimaetal2026} for the study of both). As massive stars are in most cases part of a multiple system (e.g., \citealt{Offner_etal_2023_MassiveMultipl}), the study of stripped-envelope CCSN progenitors is detrimental to observed transients (e.g., \citealt{Schulze_etal_2024_SiSCCSN} for an ultra-stripped CCSN). In some cases of binary interaction, the companion spins up the core of the CCSN progenitor. We do not consider pre-collapse core rotation in this study.
In Section \ref{sec:ExplosionTime}, we discuss the explosion time according to the JJEM and the delayed neutrino explosion mechanism.  We summarize this study in the context of the JJEM in Section \ref{sec:Summary}.

% ==========================================================
\section{Numerical Scheme}
\label{sec:NumScheme}
% ==========================================================
\setcounter{footnote}{0}

We use the one-dimensional stellar evolution code \mesa \  (Modules for Experiments in Stellar Astrophysics; version 24.08.1; \citealt{Paxton2011, Paxton2013, Paxton2015, Paxton2018, Paxton2019, Jermyn2023}) to model twelve different masses of stripped-envelope CCSNe, having solar metallicity $z = 1.42 \times 10^{-2}$ and zero-age main sequence masses ranging from $M_{\rm ZAMS} = 12 M_\odot$ and up to $M_{\rm ZAMS} = 40 M_\odot$. For each mass we simulated three scenarios of envelope stripping: a regular wind that leaves a hydrogen-rich envelope at the explosion (``full envelope''), evolution that includes an extra mass loss that removes the hydrogen but leaves helium (``H-stripped envelope''), and the removal of both hydrogen and helium (``He-stripped envelope'').\footnote{Our final models, inlists and added schemes are available online in a Zenodo upload: \href{https://doi.org/10.5281/zenodo.19737916}{doi:10.5281/zenodo.19737916}.} Our mass grid contains $12\Mo$, $15\Mo$, $18\Mo$, $20\Mo$, $22\Mo$, $25\Mo$, $28\Mo$, $30\Mo$, $32\Mo$, $35\Mo$, $38\Mo$, and $40\Mo$. We will display two of the models in Figures \ref{fig:comp} and \ref{fig:ang1}. The plots for the remaining models are available on Zenodo.

We use the default controls in the `20M\_pre\_ms\_to\_core\_collapse' test suite for all models\footnote{
The MESA EOS is a blend of the OPAL \citep{Rogers2002}, SCVH \citep{Saumon1995}, FreeEOS \citep{Irwin2004}, HELM \citep{Timmes2000}, PC \citep{Potekhin2010}, and Skye \citep{Jermyn2021} EOSes.
Radiative opacities are primarily from OPAL \citep{Iglesias1993, Iglesias1996}, with low-temperature data from \citet{Ferguson2005} and the high-temperature, Compton-scattering-dominated regime from \citet{Poutanen2017}.  Electron conduction opacities are from \citet{Cassisi2007} and \citet{Blouin2020}.
Nuclear reaction rates are from JINA REACLIB \citep{Cyburt2010}, NACRE \citep{Angulo1999}, and additional tabulated weak reaction rates \citet{Fuller1985, Oda1994, Langanke2000}.  Screening is included via the prescription of \citet{Chugunov2007}. Thermal neutrino loss rates are from \citet{Itoh1996}.
}. 
A more detailed description, including our envelope-removal method, is available in Appendix \ref{appendix:numConsistency}. We stop the simulation when infall velocity somewhere in the star exceeds $300 \kms$ but plot the results when the infall velocity is about $100 \kms$. This is because when the infall velocity becomes comparable to, and then exceeds, the convective velocity, the assumptions underlying hydrostatic MLT become increasingly questionable. Although modern MESA prescriptions (e.g., TDC MLT) include dynamical effects, we choose to characterize convection using the stable pre-collapse values rather than the rapid infall phase.

The JJEM relies on pre-collapse angular-momentum perturbations to trigger the star's explosion by seeding instabilities above the NS (Section \ref{sec:Intro}). We use the specific angular momentum parameter that previous works used (e.g., \citealt{WangShishkinSoker2024}):
\begin{equation}
    j=v_{\rm conv}r .
    \label{eq:jParameter}
\end{equation}

Both the neutrino-driven and the JJEM require these pre-collapse convection properties. This is exemplified in \cite{Varmaetal2025}, who build synthetic convective velocity fields to initialize the delayed neutrino mechanism shock revival. We find that the angular momentum fluctuations of their velocity fields have specific angular momentum fluctuations of $>4 \times 10^{15} \cm^2 \s^{-1}$. \cite{Varmaetal2025} find in their 3D simulations large convective cells with velocities of up to $\simeq 400 \km \s^{-1}$ at $r\simeq 1500 \km$ 5 seconds before collapse, namely, an angular momentum parameter of $j \simeq 6 \times 10^{15} \cm^2 \s^{-1}$. The fluctuations these studies find, which are also required for the neutrino-driven mechanism, are large enough to trigger the JJEM. 

For the minimum value of the angular momentum parameter, we consider two values $j_{\rm min}=2.5 \times 10^{15} \cm^2 \s^{-1}$ and $j_{\rm min}= 5 \times 10^{15} \cm^2 \s^{-1}$, as in earlier studies \citep{ShishkinSoker2021, ShishkinSoker2022} who justifies these as following. The minimum angular momentum for an accretion disk around an NS is $j_{\rm NS} \simeq 2 \times 10^{16} \cm^2 \s^{-1}$. As well, the spiral-SASI amplifies the perturbations (e.g., \citealt{Soker2019_SASIjets}), so the initial perturbations might be somewhat smaller, namely, only $\simeq 10^{16} \cm^2 \s^{-1}$. Consider that 3D simulations give velocities of 3-4 times as large as 1D stellar models (e.g., \citealt{FieldsCouch_2021_3Dconv}), the condition for accretion disk to form is $j_{\rm 3d} \simeq (2-4)j \gtrsim 1 \times 10^{16} \cm^2 \s^{-1}$, or $j_{\rm min}=2.5 - 5 \times 10^{15} \cm^2 \s^{-1}$.

% ===================================================
\section{Convection and Angular momentum}
\label{sec:results_models}
% ===================================================

We simulate massive stellar models up to core-collapse, as detailed in Section \ref{sec:NumScheme}, with initial masses in the range: $M_{\rm ZAMS} = 12-40\ M_\odot$. Each initial stellar model has three different mass removal prescriptions: mass loss by the regular stellar wind that leaves a hydrogen-rich envelope (full envelope) for a Type II CCSN, extra mass removal that leaves a hydrogen-deficient envelope (H envelope removed) for a Type Ib CCSN, and an additional mass removal that strips both hydrogen and helium (H-He envelope removed) for a Type Ic CCSN. 

\begin{figure*}[b]
\begin{center}
\includegraphics[trim=0cm 0cm 0cm 0cm,clip,width=\textwidth]
{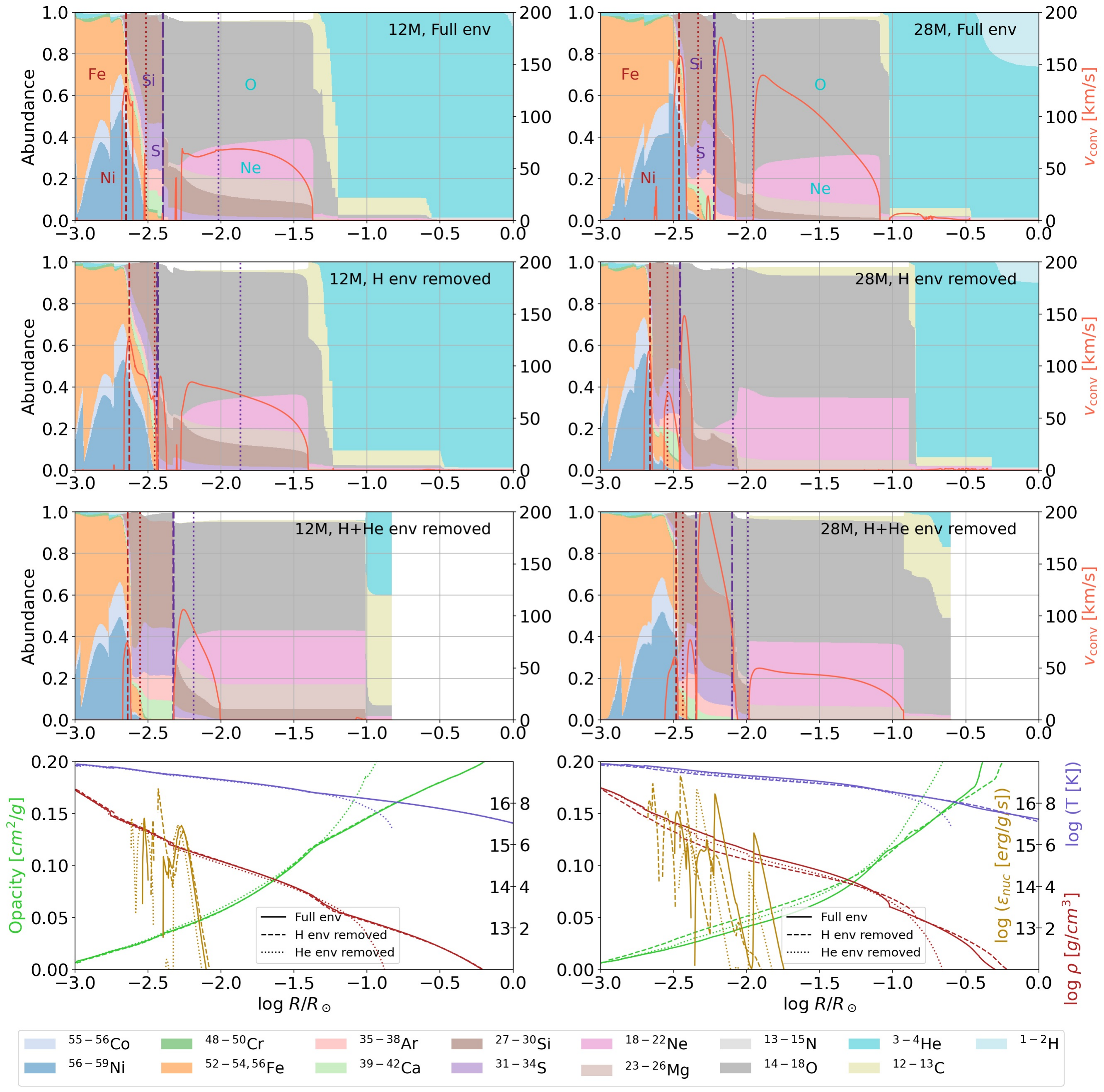} %[trim=left lower right upper]
\caption{Composition profiles of $12 \Mo$ (left column) and $28 \Mo$ (right column) models at the instance when $v^{m < 4 M_\odot}_{\rm{infall}}=100 \kms$. In the top three panels, abundances are shown as blocks of colors corresponding to the left vertical axis. Convective velocity is shown in the red line, corresponding to the right vertical axis. Vertical lines mark the Fe/Si and Si/O boundaries. The inner and outer dotted lines marks the region where $X_{\rm A>46}$ and $X_{\rm ^{28}Si+\ ^{32}S}$ are less than $0.1$, respectively; the inner and outer dashed lines mark Fe/Si boundary as defined by \mesa \  and Si/O interface where $X_{\rm ^{28}Si+^{32}S}>0.1$ and $X_{\rm ^{16}O+ ^{20}Ne + ^{24}Mg}<0.1$, respectively; the dashed-dotted purple line marks where $X_{\rm ^{28}Si+ ^{32}S} > X_{\rm ^{16}O+ ^{20}Ne +^{24}Mg}$. The lower panels show the following properties of all three models as a function of radius: opacities are shown as increasing green lines corresponding to the left vertical axis; energy generation rate per gram in units of $\erg \g^{-1} \s^{-1}$ are shown in gold lines with several peaks corresponding to the right inner vertical axis; temperature and density are shown in decreasing lines, being blue and red respectively and with density having a sharper drop, on the outer right vertical axis. Solid line corresponds to full envelope models, dashed lines to H envelope removed, and dotted line to H and He envelope removed. This figure shows strong convection at the Fe/Si interface and in the inner oxygen layer in all models, mass coordinates that correspond to the formation of an NS. The boundary lines are relevant for the discussion on explosion.}
\label{fig:comp}
\end{center}
\end{figure*}

The presence of convective zones, which are crucial for the JJEM, in the pre-collapse core, is strongly related to the location of nuclear burning and the composition. For that, we start by presenting the composition, nuclear burning, and other stellar properties, alongside the convective velocity in the pre-collapse core. 
In Figure~\ref{fig:comp} we present these for two models, $M_{\rm ZAMS} = 12 M_\odot$ and $M_{\rm ZAMS} = 28 M_\odot$, for the three envelope compositions, all at the onset of core collapse (totaling six cases). Note that we truncate the plots at $ r=1R_\odot$. Therefore, for the full and H-removed envelopes, we do not include the envelope. We plot as a function of radius, as it is more relevant to the specific angular momentum parameter $j$ (Equation \ref{eq:jParameter}) and the collapse times.
% FFFFFFFFFFFFFFFFFFFFFFFFFFFFFFFFFFFFFFFFFFFFFFFFFFFFFFFFFFFFF
% FFFFFFFFFFFFFFFFFFFFFFFFFFFFFFFFFFFFFFFFFFFFFFFFFFFFFFFFFFFFF

We denote several core boundaries with vertical lines. Generally, from the innermost to the outermost: the darker red dashed line marks the iron core boundary as defined by \mesa, where $X_{\rm A>46}>0.1$ and $X_{\rm Si+S}<0.1$; the darker red dotted line marks the point where $X_{\rm A>46} > 0.1$; the purple dashed line marks where $X_{\rm ^{28}Si+\ ^{32}S}>0.1$ and $X_{\rm \rm ^{16}O+ ^{20}Ne + ^{24}Mg}<0.1$; the purple dashed-dotted line where $X_{\rm Si} + X_{\rm S} > X_{\rm O} + X_{\rm Ne}$ (`majority vote'); the purple dotted line marks where $X_{\rm ^{28}Si+\ ^{32}S} > 0.1$. The mass coordinates where $X_{\rm A>46}>0.1$ or $X_{\rm Si+S}>0.1$ (dotted lines) are shown as an indicator of how much mixing is present at the edge of the core and are often indicative of active burning, and hence convective layers. In the composition panels of Figure~\ref{fig:comp}, the regions between the different core boundary definitions accompany composition gradients and facilitate nuclear burning and convection. This is also relevant to the discussion of the explosion time (Section \ref{sec:ExplosionTime}), as the accretion of these interface layers onto the compact remnant can fuel an explosion within the realm of the JJEM.

We note that for lower mass massive stars, which in our study are the $M_{\rm ZAMS} = 12, 15, 18 M_\odot$ models, small differences in initial mass, convection parameters, and other settings that affect burning and mixing processes can influence shell burning episodes and lead to vastly different iron core masses and Si burning (e.g., \citealt{Sukhbold_etal_2018_HighResCCSNprog}). In Appendix \ref{appendix:numConsistency}, we compare three models with very close initial masses. 

We observe that generally the outermost convective region in the core (the ONe layer) exhibits higher convective velocities for models where the envelope is removed. This shows that, despite the stripped envelope models having a smaller radius at the convective region, the specific angular momentum is compensated by the increase in convective velocity.

Even though there is no distinct relation between the amount of removed envelope mass and the convective velocity in the outer core, we can try to explain this phenomenon. As seen in the bottommost panels of Fig.~\ref{fig:comp}, the lower overall mass results in lower densities and temperatures in the outer core layers of the envelope-stripped cases, thereby slowing the burning.
This leads, in most cases, to less complete burning of the Si+S layers at the onset of collapse, which retains high burning (and energy generation) rates, fueling vigorous convection up to collapse, as opposed to the full-envelope cases where burning has already completed or is located further out. 
Additionally, the temperature and composition gradients near the core's edge are generally steeper. The combined effect of a steeper temperature and composition gradient causes higher opacity in the outer regions of the core in stripped-envelope models. The ONe region of the stripped-envelope models at the edge of the core is therefore more prone to convection. Therefore, the same amount of energy generated by nuclear burning leads to higher convective velocity in the outer core of stripped-envelope models due to higher opacities and lower densities.

We find that in all simulated models except for the $M_{\rm ZAMS}=40 M_\odot$ with H+He removed envelope, there is at least one layer with $j>2 \times 10^{15} \cm^2 \s^{-1}$ inside the mass coordinate that forms an NS. The relevant zones for NS formation are those inward to the maximum baryonic mass for NS formation (above that mass, the NS collapses to a black hole). The maximum NS gravitational mass possible is $M_{\rm NS} \simeq 2.25\ M_\odot$ (e.g., \citealt{Fan_etal_2024_MaxNSmass}). The corresponding baryonic mass, according to the baryonic to NS gravitational mass relation from \cite{Gao_etal_2020_GravBaryonMassRel}, is $M_{\rm NS,b}^{\rm max} \simeq 2.65\ M_\odot$. In the case of $M=40 M_\odot$ with H+He removed envelope there is a zone at $m \simeq 2 M_\odot$ with $j(m)>10^{15} \cm^2 \s^{-1}$. 

We present our results, specifically looking for zones with large angular momentum parameter at mass coordinates $< M_{\rm NS,b}^{\rm max}$. In this section, we present two models, of $M_{\rm ZAMS}=12M_\odot$ and $M_{\rm ZAMS}=28M_\odot$.

\begin{figure*}[b]
\begin{center}
\includegraphics[trim=0cm 0cm 0cm 0cm,width=\textwidth]{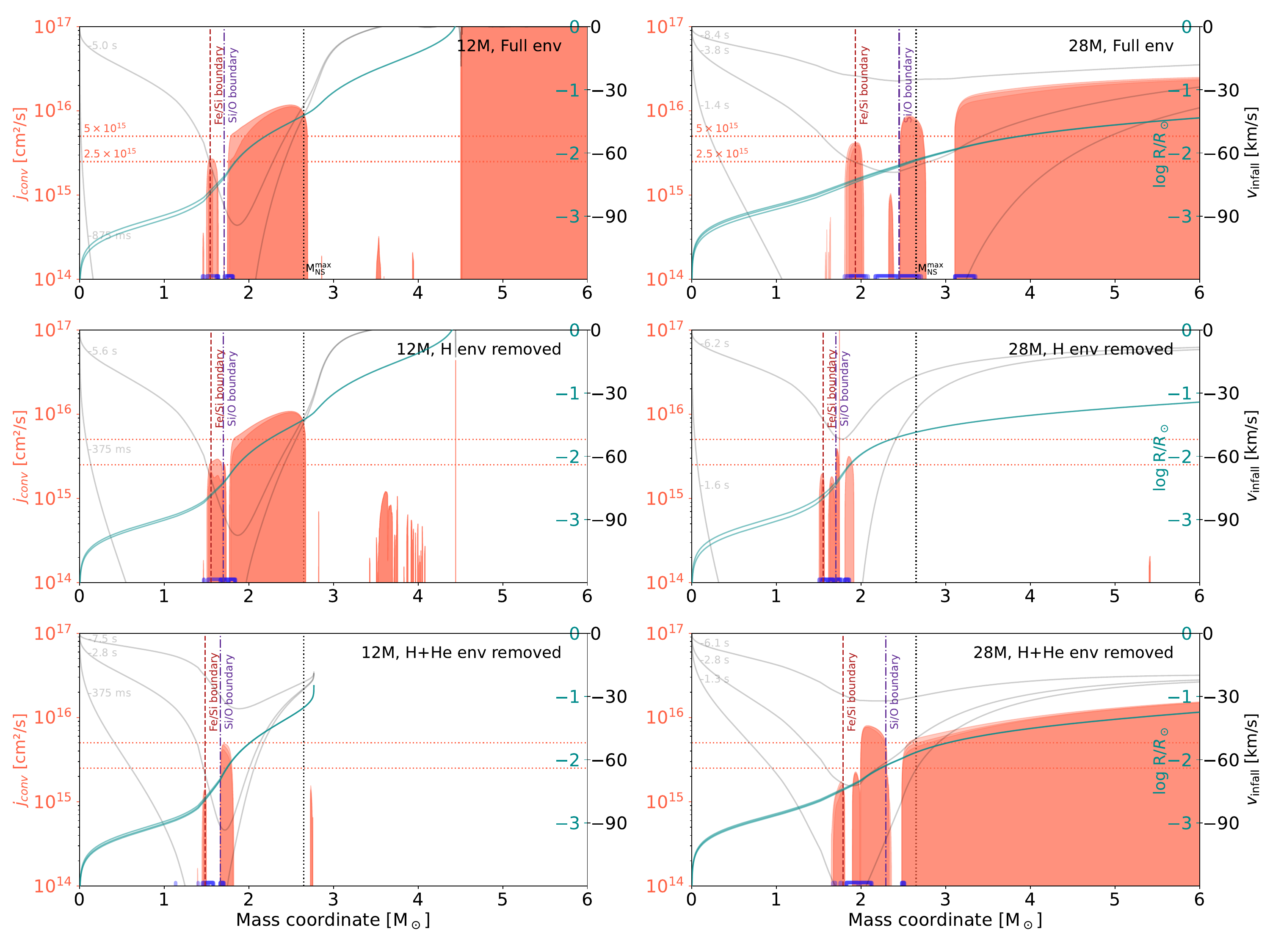} %[trim=left lower right upper]
\caption{Angular momentum parameter profiles, $j(m)$ at several times shown by red area intensity, starting from $\simeq 10 \s$ before the time when $v^{m < 4 M_\odot}_{\rm{infall}}=100 \kms$ for $12 M_\odot$ (left) and $28 M_\odot$ (right) models. The scale of $j(m)$ is on the left axis. A turquoise line represents the radius as a function of mass, with the scale on the inner right vertical axis. Black lines show the infall velocity $v_{\rm infall}(m)$ at several times, the upper line at the earliest time and the lower line at the last time, with the scale on the outer right vertical axis. The time of each snapshot is noted above the $v_{\rm infall}(m)$ line, showing $t_{\rm v_{infall} = 300\kms}-t_{\rm snapshot}$. The Fe/Si and Si/O boundaries are marked as vertical dashed lines. We denote regions with active burning ($\epsilon_{\rm nuc}>10^{14}\ \erg/\rm s$) as a blue line at the bottom of the panels, where again darker shades indicate more models with active burning at that mass coordinate. We draw two horizontal red dotted lines at $j_{\rm conv} = 2.5 \times 10^{15} \cm^2 \s^{-1}$ and $j_{\rm conv} = 5 \times 10^{15} \cm^2 \s^{-1}$, the approximate criteria for JJEM that \cite{ShishkinSoker2022} suggested. A dotted vertical line denotes the mass coordinate (baryon mass) of $M_{\rm NS,b}^{\rm max} =2.65\ M_\odot$, corresponding to the upper limit for neutron stars. 
Most models obtain specific angular momentum values larger than  $j_{\rm conv} = 5 \times 10^{15} \rm \space cm^2 \space s^{-1}$ below $m_{\rm NS}^{\rm max}$. This means that, for the envelope-removed cases, the oxygen-burning region below this limit supplies the large angular-momentum fluctuation to be accreted by the compact remnant as collapse occurs.
}
\label{fig:ang1}
\end{center}
\end{figure*}

In Figure~\ref{fig:ang1}, we present in red areas the angular momentum parameter profiles for the last $\simeq 10 \s$ before the onset of collapse. We define the onset of collapse here as the first time at which the maximum value of the infall velocity within $m < 4 \Mo$ exceeds $100 \kms$. Black lines are the infall velocities as a function of mass, $v_{\rm infall}$; the upper line is the first one, and the lower line is the last infall velocity profile we show. The turquoise lines show the radius $r(m)$ as a function of mass at these times. As the star collapses, the turquoise line representing the radius decreases over time. 
% FFFFFFFFFFFFFFFFFFFFFFFFFFFFFFFFFFFFFFFFFFFFFFFFFFFFFF
% FFFFFFFFFFFFFFFFFFFFFFFFFFFFFFFFFFFFFFFFFFFFFFFFFFFFFF

The convective zone edges and their convective velocities change between adjacent time steps in 1D simulations, due to numerical effects related to mass resolution and the mixing length theory, particularly during the collapse \citep[e.g.,][]{ShishkinSoker2021}. 3D simulations show more vigorous convection (faster motion), and smoother behavior, such as less overshooting and more consistent convective zones \citep[e.g., Figure 10 in][]{Fields_Couch_2021}.
To correctly capture the convective behavior at the beginning of the collapse, we plot profiles at several times in Figure~\ref{fig:ang1} rather than just the last one.
Each profile is plotted in a faint transparent color on top of the previous profiles. Therefore, if convection persists during this $\simeq 10 \s$ before collapse, the angular momentum parameter, $j = v_{\rm conv}r$, will be seen in a more opaque shade. A similar logic applies for the regions with active burning ($\epsilon_{\rm nuc} > 10^{14} \erg \g^{-1} \s^{-1}$). A more opaque blue line indicates that the region is actively burning for a larger portion of this time frame. There is no indicator of time progression in Figure \ref{fig:ang1} due to the stochastic nature of convection and the location of burning.
 
Some earlier studies of the JJEM (e.g., \citealt{ShishkinSoker2021}) have applied the $j(m)$ parameter to set the criterion for a convective layer to seed perturbations that lead to the formation of intermittent accretion disks, which launch the exploding jets. However, the JJEM has not yet determined the critical value of $j(m)$ for its operation. 
Following \cite{ShishkinSoker2022}, we mark two limits, $j=2.5 \times 10^{15} \cm^2 \s^{-1}$ and $j=5 \times 10^{15} \cm^2 \s^{-1}$, with dotted horizontal lines in Figure \ref{fig:ang1},  and Figure \ref{figApp:ang_15pm} in the Appendix.

The main conclusion from Figures~\ref{fig:ang1} and the models present on Zenodo, is that in massive envelope-removed stars, the inner convective zones with large enough $j(m)$ to facilitate the JJEM tend to move to lower mass coordinates. This can be either the region fueled by oxygen burning (as in $M_{\rm ZAMS}=30 M_\odot$ and $35 M_\odot$) or silicon burning (as in $M_{\rm ZAMS}=25 M_\odot$).

We note that the convective zones vary from one model to another. For example, the H-striped $28 M_\odot$ model (middle right of Figure \ref{fig:ang1}) has thin convective zones in the inner core, while in the cases of this initial model without envelope removal or with helium removal, the convective zones are more extended (upper and lower panels on the right of Figure \ref{fig:ang1}). The thick convective zones on the H-striped model are sufficient to facilitate the JJEM. The pre-collapse envelope convection is robust.

% ===================================================
\section{On the explosion time}
\label{sec:ExplosionTime}
% ===================================================
% ===================================
\subsection{General considerations of the JJEM}
\label{subsec:General}
% ===================================

In the neutrino-driven mechanism, the shock revival is not a spherical process but highly non-spherical, with protruding fingers that expand outward in narrow zones (e.g., \citealt{WongwathanaratEtal2015_shockrevival}). 
In many cases, we argue below that, even if the neutrino mechanism could have worked by itself, jets would have revived the shock first and supplied more energy.

If the jets carry a fraction $\epsilon$ of the mass accreted through the intermittent accretion disks, and the jets have a velocity of $v_j \simeq 10^5 \km \s^{-1}$, the energy the two jets carry is (\citealt{PapishSoker2011})
\begin{equation}
    E_{\rm 2j} \simeq 10^{51} \left(\frac{\epsilon}{0.1}\right)\left(\frac{M_{\rm CL}}{0.1 M_\odot}\right) \left(\frac{v_j}{10^5 \km \s^{-1}}\right)^2 \erg. 
\end{equation}
$M_{\rm CL}$ is the mass in the convective layer that forms the accretion disk; its lower bound is roughly $\simeq 0.08\Mo$ in the silicon burning layer, as in the full envelope $M_{\rm ZAMS}=15
M_\odot$ (Zenodo) and H+He stripped $M_{\rm ZAMS}=12 M_\odot$ (Figure~\ref{fig:ang1}). The upper value for $M_{\rm CL}$ in the inner core can be as high as $\simeq1\Mo$ (Full envelope $M_{\rm ZAMS}=12 M_\odot$, Figure~\ref{fig:ang1}). Previous studies have also estimated the required $M_{\rm CL}$ for explosion as $\simeq 0.1\Mo$ (e.g., \citealt{ShishkinSoker2022}).
We scale for the fraction of the accreted mass that the jets carry by $\epsilon =0.1$ as in previous JJEM papers (see review in \citealt{Soker2025Learning}), and based this ratio on other astrophysical types of objects launching jets, such as young stellar objects (e.g., \citealt{NisiniEtal2018_epsilonYSO}).
Even if the mass in the first convective layer is smaller, the jets will at least penetrate the stalled shock and revive it along their paths. Once the jets break out, neutrino heating supplies more energy along that direction and boosts the jets' energy \citep{Soker2022_Boosting}. 

If the jets of the inner convective layer do not manage to explode the entire core, the outer convective layers that are accreted later will form more intermittent accretion disks that will launch more jets until the entire core is exploded. In any case, these early jets will revive the stalled shock at specific locations before the neutrino-driven mechanism does.

According to the JJEM, the explosion begins when the newly born NS accretes the convective layers from the core. These are generally the regions of and near the Fe/Si interface (silicon burning zone) and the Si/O interface (oxygen burning zone). We comment in this section on the timescales of the explosion, as described by the JJEM and the neutrino-driven mechanism. 

On average, each convective layer launches several pairs of jets \citep{GilkisSoker2014}. During the accretion period of the convective layer, several pairs of jets can accumulate to provide most of the explosion energy. The JJEM operates in a negative feedback mechanism \citep{PapishSoker2011}, and the explosion energy is several times the binding energy, with large variations as the jet launching process is stochastic (e.g., review by \citealt{Soker2024Rev}). When the jets have enough energy to explode the star in all directions, accretion onto the newly born NS ceases. The outcome is that the explosion energy is typically several times the binding energy of the ejecta. The reason is that the explosion is not 100 percent efficient, in the sense that, along many directions, the jets accelerate material to velocities above escape velocity, forming lobes and ears. In addition, even if there is already enough energy in the jets to explode the star, some material from the core is already accreted, leading to late jets that add more energy. \cite{ShishkinSoker2023} discussed the binding energy implications in envelope-removed core-collapse progenitors, based on the energetic estimates of the JJEM from \cite{GilkisSokerPapish2016}, and found the conditions for explosion met in a wide range of progenitor models.

% ===================================
\subsection{Onset time of the explosion}
\label{subsec:Onset}
% ===================================

In a recent paper, \cite{Bocciolietal2025Compact} compared the time of accretion of the Si/O interface with the time the revived shock reaches a radius of $500 \km$ in their one-dimensional (1D) simulations of the neutrino-driven mechanism. They get some successful shock revival with their 1D simulations, but many models failed to explode. Those models with solar metallicity that do explode yield explosion energies of $0 \lesssim E_{\rm exp} (1D) \lesssim 10^{51} \erg$ \citep{Bocciolietal2025Si}.  
Here, we compare the time at which the stalled shock begins to re-expand with the time of Si/O interface accretion to elaborate on the relationship between the two explosion mechanisms. 

\begin{figure}[]
	\begin{center}
\includegraphics[trim=0.0cm 18.5cm 0.0cm 0.0cm ,clip, scale=0.54]{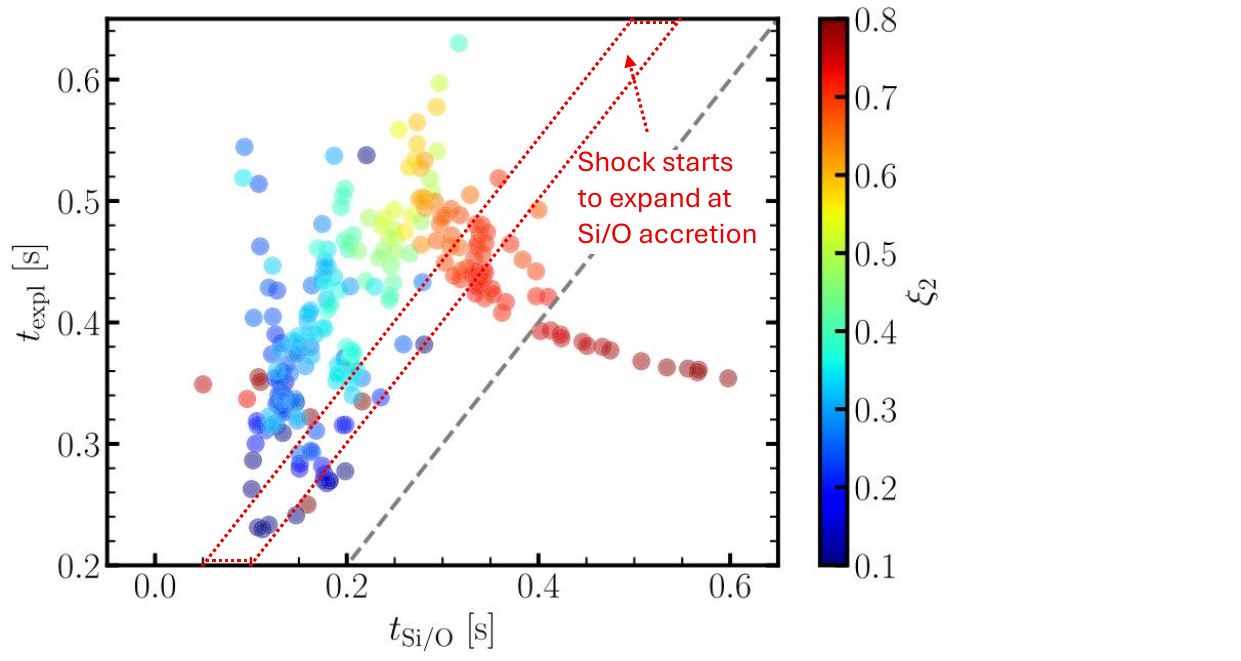} 
\caption{A figure adapted from \cite{Bocciolietal2025Compact} where they show the explosion time $t_{\rm expl}$, defined as the time the revived shock reaches a radius of $500 \km$, as a function of the time the Si/O interface from the core is accreted onto the shock. Times are measured from the shock bounce. Their dashed line shows $t_{\rm expl}=t_{\rm Si/O}$. The revived shock reaches the radius of $500 \km$ after $\Delta t \simeq 0.1-0.15 \s$. We added the zone enclosed by the red-dotted lines to indicate the equality of Si/O accretion and the start of the shock re-expansion. In most of their models, the shock starts to expand after the Si/O is accreted. Only in those with large compactness $\xi_2 \equiv 2000 \km /R(2)  \gtrsim 0.6$ the explosion is much shorter than the Si/O accretion time; $R(2)$ is the radius enclosing $2M_\odot$ baryonic mass in the pre-collapse core.  
}
\label{Fig:SiOaccretion}
\end{center}
\end{figure}

In Figure \ref{Fig:SiOaccretion} adapted from \cite{Bocciolietal2025Compact}, they present the time the shock reaches a radius of $500 \km$, $t_{\rm expl}$, as a function of the time the Si/O is accreted onto the shock for their models that explode. It takes the shock $\Delta t \simeq 0.1-0.15 \s$ to reach $500 \km$ from the moment it starts to rapidly re-expand. Namely, the revival time is $t_{\rm rexp} \simeq t_{\rm expl} - \Delta t$. We added the zone between the two diagonal red-dotted lines to indicate where $t_{\rm rexp} \simeq t_{\rm Si/O}$. The red-colored models have large compactness values, $\xi_2 \gtrsim 0.6$, which correspond mainly to massive progenitors, $M_{\rm ZAMS} \gtrsim 22 M_\odot$ (e.g., \citealt{Sukhbold_etal_2016}), and thus represent rare CCSNe.

\begin{figure*}[]
\begin{center}
\includegraphics[trim=0.5cm 0.0cm 0.0cm 0cm,width=\textwidth]{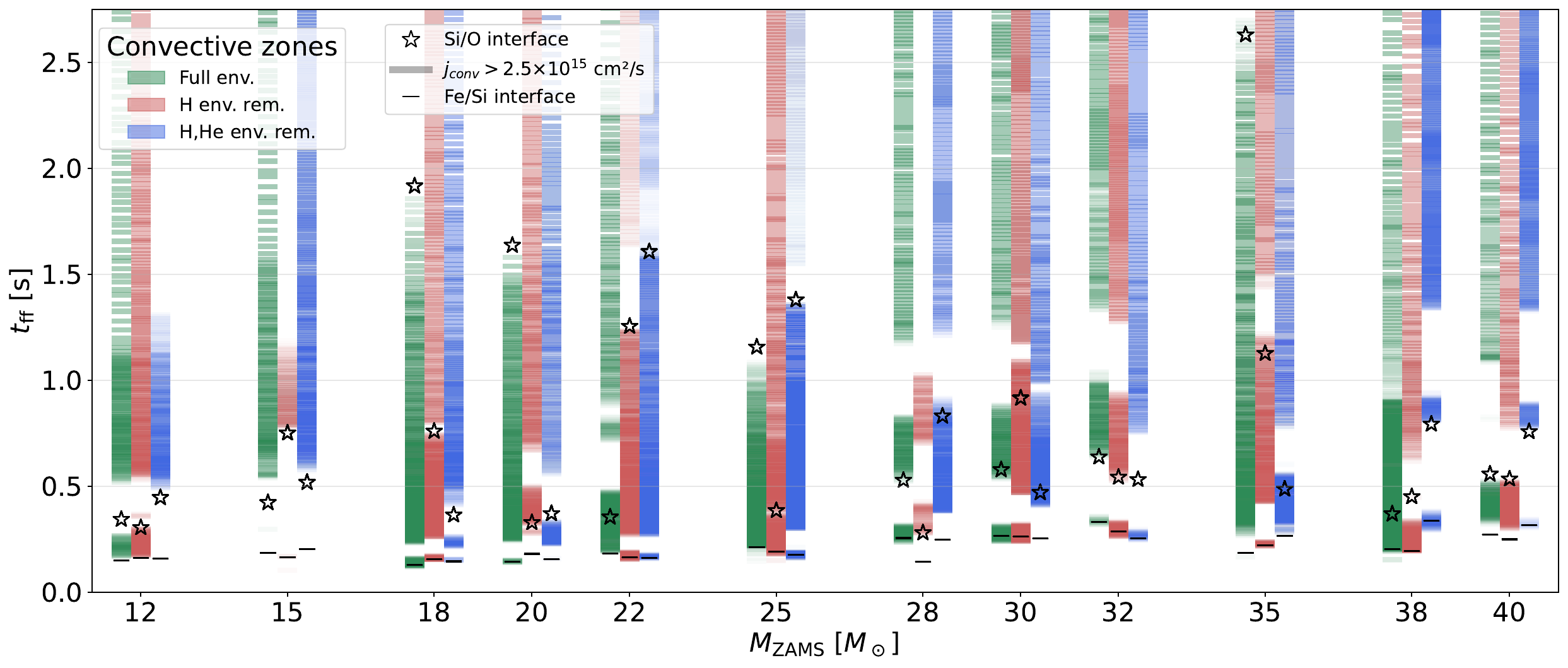} %[trim=left lower right upper]
\caption{Free fall time from the Fe/Si and Si/O boundaries when at least one mass shell at $m<4 M_\odot$ has a infall velocity of $v_{\rm infall} > 100 \kms$, i.e., at the onset of core collapse. In each column set, we mark by color the free-fall time of convective regions where $j_{\rm conv}>2.5\times10^{15} \cm^2 \s^{-1}$ (otherwise it is white). For each initial mass ($M_{\rm ZAMS}$), we display three columns corresponding to each envelope possibility; Green color represents models with a full envelope, red represents models with a hydrogen stripped envelope, and blue represents models with both a hydrogen and helium envelope removed. 
We denote the interface boundaries to indicate the relation of the vigorous convective zones (opaque colors) to the inner core structure; Short horizontal lines represent the place of Fe/Si boundary as defined by \textsc{mesa}, and star markers represent the Si/O interface where $X_{\rm Si} + X_{\rm S} > X_{\rm O} + X_{\rm Ne}$.
We examine several times just before collapse; darker colors signify that at most or all times there was convection for that free-fall time point, while a faint color indicates that only at one or a few times that we examine before collapse, the specific location had convection. We plot all profiles from the onset of collapse to simulation termination.
For most models, a vigorous (with $j_{\rm conv}>2.5\times10^{15} \cm^2 \s^{-1}$), consistent (in the figure: opaque) convective zone is located between Fe/Si boundary and the Si/O interface - that will fuel the JJEM prior to the accretion of Si/O interface and the potential shock revival of the delayed-neutrino mechanism.
The explosion time according to the JJEM requires additional delay time of $\simeq 0.02 \s$ as the material behind the stalled shock slows down substantially.}
\label{fig:Apptff}
\end{center}
\end{figure*}

Figure \ref{Fig:SiOaccretion} shows that in the majority of the models that \cite{Bocciolietal2023} study, the accretion of the Si/O interface onto the shock occurs before shock revival. The Fe/Si interface, which also has a convective zone (e.g., Figures~\ref{fig:ang1},  and Figure \ref{figApp:ang_15pm} in the Appendix), resides further inside the pre-collapse core and is accreted before the Si/O interface. To illustrate the crude timescales, we present in Figure \ref{fig:Apptff} the free-fall times from the Fe/Si and Si/O boundaries in our pre-collapse stellar models, three models for each ZAMS mass. The Fe/Si interface is accreted $\simeq 0.1-0.3 \s$ before the Si/O interface in most models.  
Figures \ref{Fig:SiOaccretion} and \ref{fig:Apptff} show that in the vast majority of CCSNe, the Fe/Si is accreted before the shock revival. This interface features a convective zone that initiates the JJEM process, which can last from a second to several seconds.  
% FFFFFFFFFFFFFFFFFFFFFFFFFFFFFFFFFFFFFFFFFF
% FFFFFFFFFFFFFFFFFFFFFFFFFFFFFFFFFFFFFFFFFff
% FFFFFFFFFFFFFFFFFFFFFFFFFFFFFFFFFFFFFFFFFFFFFFFFFFFFFFFFFFFFF
% FFFFFFFFFFFFFFFFFFFFFFFFFFFFFFFFFFFFFFFFFFFFFFFFFFFFFFFFFFFFF

As in Figure~\ref{fig:ang1}, we present in Figure~\ref{fig:Apptff} the convective layers at several times (about 2 to 10 times) starting from about $10 \s$ before, and up to the profile where infall velocity reaches $100 \kms$. The reasoning is that some convective regions may abruptly appear and disappear due to numerical effects (e.g., limited resolution and the use of the mixing-length theory during a collapse), and it is more robust to quote stable convective zones that persist over a period of time (See description of Figure \ref{fig:ang1} in Section \ref{sec:results_models}).

We note that different definitions for the interface layer location (See Figure \ref{fig:comp}) will slightly change the free-fall time but not change the overall conclusion.

Three comments are in place here. 
\newline
 (1) We took the explosion of the JJEM to be at the time of the accretion of the convective layer. The times of the accretion of the material that \cite{Bocciolietal2025Si} give are the accretion onto the shock. The time the material spends behind the stalled shock is about the mass in the gain region divided by the accretion rate $M_{\rm gain} / \dot M_{\rm acc}$. From the simulations by \cite{Mulleretal2017}, we estimate this time scale to be $\simeq 0.02 \s$ (dividing the mass in the gain region they give in their Figure 11, $\simeq 0.01 M_\odot$ by the accretion rate from their Figure 3, $\simeq 0.5 M_\odot \s^{-1}$).  This time includes the time the material goes through the unstable zone that amplifies flow perturbations, like angular momentum fluctuations. For comparison, the free-fall time from $150 \km$ to the center is only $\simeq 0.005 \s$.    
We note the following two limitations. (a) The simulation-inferred parameters for the mass in the gain region and for the accretion rate onto the proto-NS in \cite{Mulleretal2017} are strictly valid only for the $18 M_\odot$ progenitor they simulated. (b) The time the material spends behind the stalled shock is likely longer for low-compactness stars, which we focus on in this work; these are lower-mass stars with a shallower mass gradient at the Si/O interface and therefore a lower mass accretion rate onto the proto-NS.
 Accretion from the accretion disk onto the NS will begin immediately as the disk forms. Other processes, like accretion disk formation, will occur on a dynamical timescale around the NS, which is $\simeq 0.001-0.005 \s$. This is about the Keplerian time, as the disk material orbits once, collides with itself, and the low-angular-momentum part of the gas gets immediately accreted (e.g., \citealt{Soker2026Long}). 
For example, the Keplerian time of the accretion disk is 
\begin{equation}
   \tau_{\rm d} = 0.0018 \left(\frac{R}{25 \km}\right)^{3/2}\left(\frac{M_{\rm NS}}{1.4 M_\odot}\right)^{-1/2} \s 
\end{equation}
 Its build-up time can be a few times this timescale.
 Overall, the jet launching might start $\simeq 0.025 \s$ after the accretion of the convective layer onto the shock. The comparison with the explosion time requires shifting the dotted-red line in Figure \ref{Fig:SiOaccretion} by $0.025 \s$ up to align with the neutrino-driven results.  On the other hand, in most models the Fe/Si layer is convective and is accreted $\gtrsim 0.1 \s$ before the Si/O layer.  
 \newline
 (2) Figure \ref{Fig:SiOaccretion} shows that stellar cores with large compactness $\xi_2 \equiv 2000 \km /R(2) \gtrsim 0.6$ explode early, before the Si/O accretion time; $R(2)$ is the radius enclosing $2M_\odot$ baryonic mass in the pre-collapse core.  
 These models correspond to massive stars and are relatively rare. More importantly, the 1D models which were used for these calculations show the explosion energies of these models to be $>2 \times 10^{51} \erg$ (e.g., \citealt{Bocciolietal2025Si}). However, 3D simulations show that the neutrino-driven mechanism does not usually reach these explosion energies; if it does, the remnant is a black hole (e.g., \citealt{Burrowsetal2024Let}). 
 %In this study, we are not considering black hole remnants. The physics of these high-compactness models requires further study. 
 \newline
(3) Each convective layer, according to the JJEM, gives rise to a few to several jet-launching episodes. These fluctuations arise from convective cells within the convective layer. \cite{Soker2025Learning} estimated, based on observations, the duration of jet-activity episodes to be in the range of $\tau_{\rm 2j} \simeq 0.01-0.3 \s$. Such activity lifetimes span $\simeq 5-100$ Keplerian orbits of the disk at $r=25 \km$, and $\simeq 2-50$ orbits of disk material at $r=50 \km$. Short-lived accretion episodes cannot support large accretion disks. \cite{Soker2018arXiv} analyzed simulations by \cite{Mulleretal2017} and find the angular momentum fluctuations lifetimes to be $\simeq 0.05 - 0.1 \s$. \cite{Soker2026Long} suggested that angular momentum transport by viscosity extends the lifetime of some intermittent accretion disks even when material with zero average angular momentum is accreted. This explains the lifetime of the intermittent accretion disks that can be up to $\simeq 0.3 \s$. Namely, in some cases, the disk continues to live even after the accretion is from a non-convective layer. Overall, the accretion of one convective layer will supply $\simeq 3-10$ pairs of jets that explode the star (e.g., \citealt{Soker2025Learning, Soker2026Long}). 
The typical mass in the convective zones inner to the Si/O interface is $\simeq 0.02-0.1M_\odot$ (e.g., Figure \ref{fig:ang1}; \citealt{ShishkinSoker2022}). If the 3-10 pairs of jets carry $\simeq 0.2$ of the accretion gravitational energy onto an NS of mass $1.4 M_\odot$ and a radius $20 \km$, the jets' total energy is  $E_{\rm pairs} \simeq 7 \times 10^{50} - 3.5 \times 10^{51} \erg$. Not all accretion gas from one layer will go through an accretion disk. However, \cite{Soker2026Long} showed recently that the accretion disk can live longer, and 3D simulations show more vigorous convection than what 1D models give. Therefore, taking the jets to carry $\simeq 20 \%$ of the accretion energy of the convective layer is reasonable. 
We note the following limitations. (a) The JJEM theory is not yet complete enough to allow for a rigorous prediction of jet launching, of the number of jet pairs per reference mass of accreted convective material, and of the conversion efficiency of the accretion energy into kinetic energy of the jet pairs for each stellar progenitor. These values come mainly from comparison with observed CCSNRs.   (b) The results we provide are preliminary analytical estimates based on previous works (e.g., \citealt{Soker2025Learning}) and a fiducial test case (e.g., a $1.4 M_\odot$ proto-NS inflated to $20 \km$ radius before contraction, the launching of 3-10 jet pairs, and the $20 \%$ conversion efficiency of accretion energy into kinetic jets' energy); we assume these to apply to all stellar progenitors. The neutrinosphere might be larger, but we are referring to the dynamical effect of accretion where material falls to radii deeper than the neutrinosphere (e.g., \citealt{GuiletFernandez2014}).
(c) The numerical values should be more specifically tailored to the stellar progenitor properties when detailed simulations of the JJEM become available.

In addition, neutrino heating can boost the jets' energy \citep{Soker2022nu}.  Overall, the convective material within the Si/O layer can form jets that supply the entire explosion energy in most CCSNe, and trigger it in the rest. Additional convective layers can add more jets and energy if accretion onto the newly born NS continues, as we show in the next section. 

% ===================================
\subsection{On the JJEM explosion energy}
\label{subsec:Energy}
% ===================================

\begin{figure*}[]
\centering
\includegraphics[trim=0.0cm 0.0cm 0.0cm 0cm,clip, scale=0.50,width=0.45\textwidth]{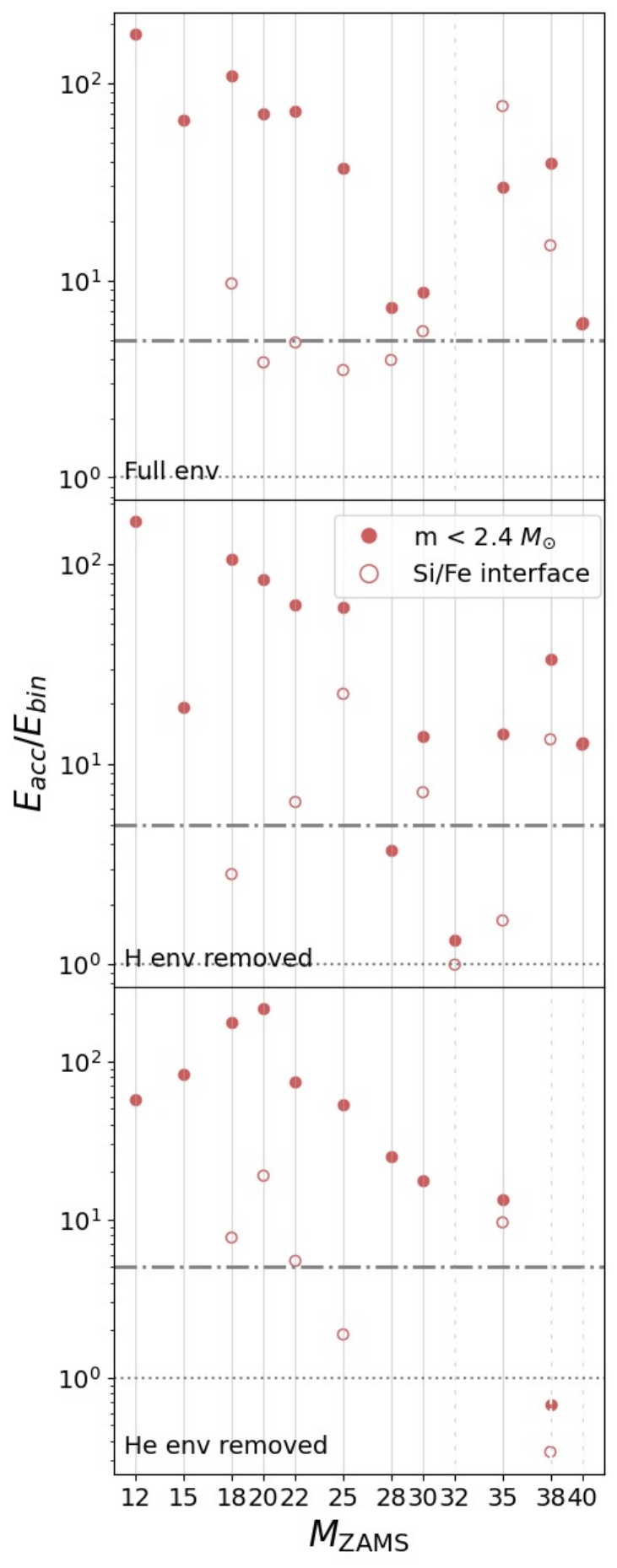} 
\caption{Ratio of accretion energy from convective layers with $j>2.5 \times 10^{15} \cm^2 \s^{-1}$ to the binding energy of mass residing above $m = 1.6\ M_{\odot}$. The open circles present ${E_{\rm acc}}/{E_{\rm bin}}$ for accretion of the innermost convective region, where and near where Si burns to Fe. Filled dots present the ratio when all convective regions within $m<2.4 \Mo$ are accreted. Each vertical line marks a model; the lack of circles on a vertical line means the lack of convective layers with $j>2.5 \times 10^{15} \cm^2 \s^{-1}$ within the radius criteria. In most models, there is enough accretion energy in the convective regions to explode the star even if the jets carry only $\simeq 20\%$ of this energy. The thick dot-dash horizontal line marks this criterion, namely, ${E_{\rm acc}}/{E_{\rm bin}} = 5$; the dotted horizontal line marks where ${E_{\rm acc}}/{E_{\rm bin}} = 1$, below which $E_{\rm bin}$ would be too high for the non-rotating star to explode by JJEM. 
}
\label{Fig:Binding}
\end{figure*}

More quantitatively, we present in Figure \ref{Fig:Binding} the ratio of the accretion energy of the convective layers with high angular momentum parameter, $j > 2.5 \times 10^{15} \cm^2 \s^{-1}$ (Equation \ref{eq:jParameter}), to the binding energy of the material residing above $m=1.6 M_\odot$, calculated from integrating the `total\_energy' as given by \textsc{MESA}. This is the approximated mass that is needed to form an NS of $1.4 \Mo$ gravitational mass, so a successful explosion needs to eject the mass above $1.6 \Mo$. The accretion energy we take is scaled as follows 
\begin{equation}
E_{\rm acc}= \frac{1}{2}\frac{G (1.4 M_\odot) M_{j>2.5\times 10^{15}}}{20 \km}, 
    \label{eq:Eacc1}
\end{equation}
where we substitute $M_{\rm NS} =1.4 M_\odot$ for the NS that accretes the mass, and $R_{\rm NS}=20 \km$ for the radius of the NS, as it is not yet relaxed to its $\simeq 12 \km$ final radius (e.g., this is the proto-NS radius at $t=0.4 \s$ as presented by \citealt{Pruetetal2005}). 
The empty circles in Figure \ref{Fig:Binding} present the accretion energy of the convective layer(s) at and near the Si/Fe interface, i.e., near and at where silicon burns to iron. The filled dots present the accretion energy of mass with $j > 2.5 \times 10^{15} \cm^2 \s^{-1}$ in all convective layers in the inner region $m<2.4 M_\odot$, i.e., the mass that will not lead to black hole formation.
% FFFFFFFFFFFFFFFFFFFFFFFFFFFFFFFFFFFFFFFFFF
% FFFFFFFFFFFFFFFFFFFFFFFFFFFFFFFFFFFFFFFFFff

We assume that the explosion condition is that $20\%$ of the accretion energy of convective material up to mass coordinate $m=1.6 M_\odot$ onto the proto-NS is larger than the gravitational binding energy of the stellar mantle above $m=1.6 M_\odot$. 
Figure \ref{Fig:Binding} shows that in most of the models the convective layer near the Si/Fe interface can supply enough energy to overcome the binding energy at $m = 1.6 \Mo$. When we include all convective layers with $m<2.4 \Mo$ with a sufficiently high angular momentum parameter, $j > 2.5 \times 10^{15} \cm^2 \s^{-1}$, we find that all models but four have enough $E_{\rm acc}$ for explosion. The $M_{\rm ZAMS}=32 M_\odot$ full envelope model does not explode with neutron star formation. It will explode by convection layers further out since there is sufficient convection in the oxygen-burning region, but will form a black hole (there are no failed CCSNe in the JJEM; e.g. \citealt{Soker2026Failed}). 
Three of the models that have low accretion energy are He-envelope-removed models with initial masses $M_{\rm ZAMS}=32,\ 38,\ 40 M_\odot$, where we expect the envelope to be stripped by a companion that spins up the star. Therefore, the main source of angular momentum in the stripped case is not accounted for in this study, and these models can have sufficient $E_{\rm acc}$ if we take binary interaction into account. Also, some of these stars, according to the JJEM, form black holes, as we explain next. 

In most cases, even if the jets carry only $20\%$ of the accretion energy, they alone can unbind the star (as said, neutrino heating adds some energy). In some of the models with stripped envelopes, the jets might not unbind the entire star, leading to black hole formation. Indeed, according to the JJEM, some stripped-envelope stars that were spun up by a companion (which also stripped their envelope) explode via a pair of constant-axis jets, but more of the core mass near the equatorial plane is accreted to form a black hole.

The new point we state here is that the convective regions, the Fe/Si interface in the majority (and possibly all) cases, and the Si/O interface, in most models, are accreted before the neutrino-driven mechanism revives the stalled shock. \textit{Namely, even when the neutrino-driven mechanism might have led to an explosion (but generally of too low energy), the JJEM starts to operate before the neutrino mechanism fully revives the stalled shock.} 
The Crab Nebula suggests this is true even in low-energy CCSNe, because the Crab Nebula is a low-energy supernova (e.g., \citealt{YangChevalier2015_crabEnergy}) that possesses a clear point-symmetric morphology that we argue only a jet-driven explosion can explain \citep{ShishkinSoker2025Crab}.

% ==========================================================
\section{Summary}
\label{sec:Summary}
% ==========================================================

The JJEM asserts that pre-core-collapse convection seeds angular momentum perturbations that form intermittent accretion disks, which launch jittering pairs of jets that explode CCSNe. The shaping of CCSNRs by two or more pairs of jets is a robust finding of the last four years (Section \ref{sec:Intro}). Using the one-dimensional stellar code \textsc{mesa} (Section \ref{sec:NumScheme}), including no magnetic fields or rotation, we examined the specific angular momentum parameter $j(m)$ (Equation \ref{eq:jParameter}) in the cores of pre-collapse stellar models, focusing on stripped-envelope stars, i.e., progenitors of SNe Ib and Ic. Earlier studies of the JJEM used this parameter as an indication of the presence of these seed perturbations.  

The main results of the simulations are the red areas in Figure \ref{fig:ang1} and other similar figures present on Zenodo. These results indicate that in almost all models, regardless of whether they have hydrogen-rich, hydrogen-stripped, or helium+hydrogen-stripped envelopes, there is at least one significant strong convective zone in the inner core. Namely, a prominent zone with $j(m) \gtrsim 2 \times 10^{15} \cm^2 \s^{-1}$ in the inner core. By inner core, we refer to the core inside the mass coordinate that is the maximum baryonic mass of a NS, $M_{\rm NS,b}^{\rm max} \simeq 2.65\ M_\odot$ (Section \ref{sec:results_models}).  Our finding is encouraging for the JJEM, although it does not show that the intermittent accretion disks indeed form. 

In this theoretical study, we examined convective motion using the mixing-length theory (MLT) as implemented in \textsc{mesa}. Three-dimensional simulations reveal that the maximum convection velocity in three dimensions is several times faster than the velocity that the MLT gives, particularly for low angular wave-numbers (e.g., \citealt{FieldsCouch_2020_3Dconv, Yadav_etal_2020_3Dconv, Yoshida_etal_2021_3Dconv, FieldsCouch_2021_3Dconv, Georgy_etal_2024_3Dconv, Andresenetal2026}). This suggests that the convective zones we simulate in 1D correspond to large-scale motion with high specific angular momentum at the onset of collapse in real three-dimensional cases. The requirement for large, fast convective cells in the pre-collapse core is common to both the neutrino-driven mechanism and the JJEM. In a recent study, \cite{Varmaetal2025} generated synthetic 3D velocity fields in the pre-collapse convective zones of cores. The convection models they construct have large scales (low spherical harmonic degree $l$) and fast velocities. 

Moreover, actual convection just before core collapse might be more vigorous than what these simulations give. In a recent study of metal distribution in the CCSN remnant Cassiopeia A, \cite{Satoetal2025} concluded that within hours before explosion, there was a strong mixing of the oxygen layer with the adjacent layers in the core. This mixing requires vigorous convection. \cite{Satoetal2025} comment that such perturbations in the collapsing core can facilitate shock revival in the framework of the neutrino-driven mechanism. We comment that such vigorous convection can seed large perturbations and lead to the formation of stochastic intermittent accretion disks that launch jittering jets that explode the star, i.e., the JJEM. Both the JJEM and the neutrino-driven mechanism require strong pre-collapse perturbations.

Studies of stripped-envelope CCSNe in the framework of the neutrino-driven mechanism predict many failed supernovae, namely, a direct collapse to a black hole of most of the star (e.g., \citealt{Zapartasetal2021}). We find that almost all our models in this study, including stripped-envelope cases, have sufficient convection to seed the JJEM and explode as CCSNe, leaving behind an NS when core rotation is negligible. According to the JJEM, black holes are formed when the pre-collapsed core is rapidly rotating (e.g., \citealt{Soker2023gap}). The reason is that, in the case of rapid pre-collapse core rotation, the accreted gas has angular momentum along the same axis as the pre-collapse rotation, with very little jittering around it. The jets are inefficient at expelling gas from the equatorial plane where the accretion disk forms. Accretion continues from the equatorial plane, transforming the remnant into a black hole while the jets explode the star along the polar directions. Namely, this is an explosion rather than a failed supernova, even when a black hole forms, and it can be energetic (e.g., \citealt{GilkisSokerPapish2016}). Jets in the cores of massive stars can inflate large cocoons (lobes; e.g., \citealt{Gottliebetal2023}) that expel mass from the equatorial plane as well. This, we speculate, can be effective for lower-mass cores, but not for massive cores with masses $\gtrsim 10 M_\odot$. In a future study, we will examine whether this might explain the peak of black hole masses around $10 M_\odot$ that gravitational wave observations of merging black holes infer (e.g., \citealt{GWTC40LVK2025}).

In Section \ref{sec:ExplosionTime}, we showed that in the majority, but not all, of stellar models where the neutrino-driven mechanism revives the stalled shock, we expect the JJEM to start operating at earlier times. This finding adds to the difficulties of the neutrino-driven mechanism to account for several observational properties of CCSNe and CCSN remnants, such as point-symmetric CCSN remnants, large explosion energies of some CCSNe, and the non-detection of `failed supernovae'; see the list of challenges of the neutrino-driven mechanism in \cite{Soker2025Learning}.  We, therefore, support the claim of earlier studies (see Section \ref{sec:Intro}) that the JJEM can independently account for all CCSNe.

Previous studies (see Section \ref{sec:Intro}) noted that the progress in establishing the JJEM as the primary explosion mechanism of CCSNe has been made in many small steps. Upon analyzing observations, the identification of a new point-symmetric CCSNR represents a small step (there have been about 20 such steps to date). At the same time, theoretically, there are smaller simulation steps, such as three-dimensional hydrodynamical simulations of shaping by jittering jets (e.g., \citealt{Braudoetal2025, Braudoetal2026, AkashiSoker2026a, AkashiSoker2026bG11}). Our study represents a small step toward demonstrating robust convection in the inner core of stripped-envelope CCSN progenitors and toward finding that the JJEM might explode the star before shock revival in the neutrino-driven mechanism.

% ===================================================
\section*{Acknowledgments}
% ===================================================
We thank an anonymous referee for very detailed comments that helped us refine and improve the presentation of our results. NS thanks the Charles Wolfson Academic Chair at the Technion for the support. 

% ======================================
% ======================================
% ======================================
%%%%%%%%%%    Use one of the below.
%%%%%%%%%%  Remove the %%%% in the two lines below to replace the list of references.
 \bibliography{bib}{}
 \bibliographystyle{aasjournal}

% ======================================
% ======================================
%%%%%%%%%%  Insert below the references for the arXiv: 
% =======================================

% ======================================
% ======================================
% ======================================

\setcounter{figure}{0}      
\renewcommand{\thesection}{\Alph{section}}
\appendix
% =========================================
\section{Appendix A: Envelope removal scheme}
\label{appendix:removalTiming}
% =========================================
\renewcommand\thefigure{A} %.\arabic{figure}

Our envelope removal scheme favors a physical criterion that is radius-dependent. Namely, we remove the hydrogen envelope at the rapid expansion when the star is nearing hydrogen depletion at the centre, specifically when the star (photosphere) reaches a radius of $R_*=800\ R_\odot$.
This is different from the default scheme used for massive stars, as implemented, for example, in the `20M\_pre\_ms\_to\_core\_collapse' test suite, which removes the hydrogen envelope at central helium depletion.

In this section, we compare the two different removal schemes in terms the stellar parameters of mass, radius, inner composition (mass, density and temperature profiles) and observables (luminosity and effective temperature; as a Hertzsprung–Russell (HR) diagram).

\begin{figure}[h]
\centering
    \includegraphics[trim=0cm 0cm 0cm 0cm, width = 0.8\textwidth]{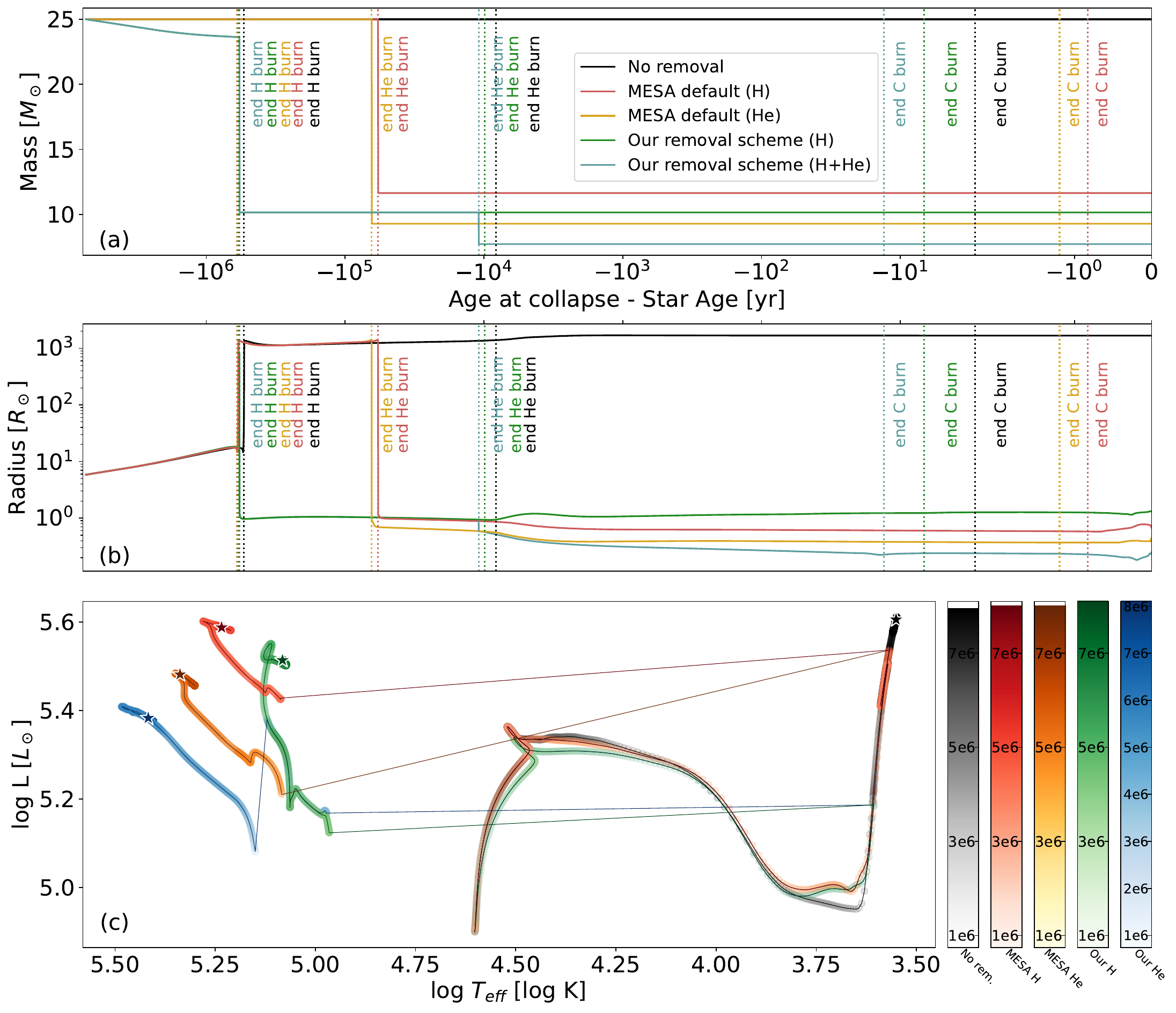}
    \includegraphics[trim=0cm 0cm 0cm 0cm, width = 0.8\textwidth]{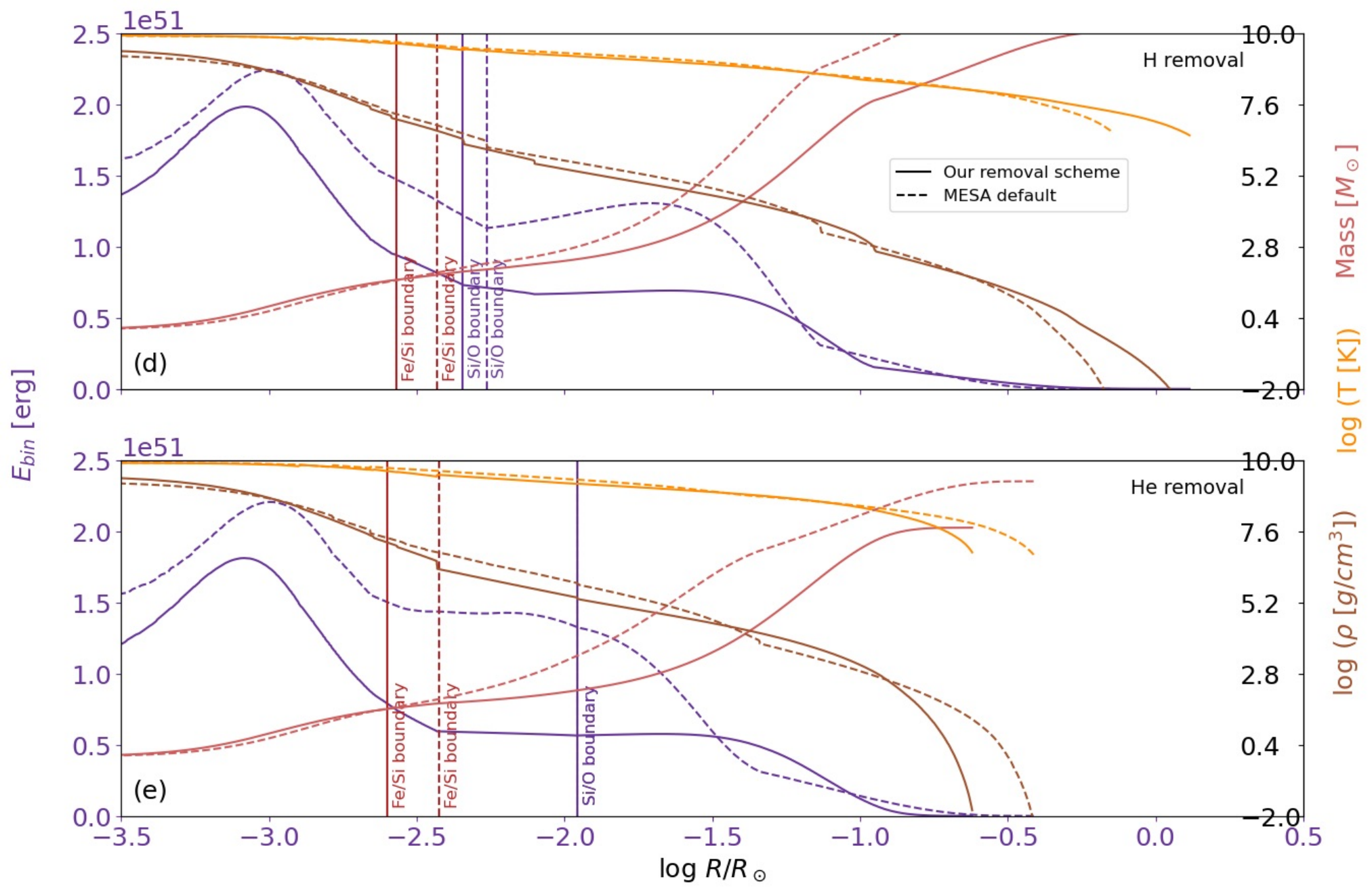}
    \label{figApp:compRem}
\caption{Envelope removal schemes comparison.
\textbf{(a–c)} Time evolution of models with different envelope-removal prescriptions and levels. Panel (a) shows total stellar mass as a function of time up to collapse, while panel (b) presents the corresponding stellar radius evolution. Panel (c) displays the full evolutionary tracks in the HR diagram for the five models with different degrees of envelope removal and numerical schemes. Colors progress from light to dark with time, as indicated by the colorbars on the right. A star marker denotes the end-point of the simulation on the HR tracks (core-collapse).
\textbf{(d–e)} Comparison between envelope-removal schemes for a $25,M_\odot$ model. Panel (d) shows the H-removal cases, while panel (e) shows the H+He-removed cases. In both panels, the binding energy is plotted in purple and scaled on the left vertical axis, while mass, temperature, and density are scaled on the right vertical axis. All profiles are taken at the first profile to satisfy $v^{m < 4,M_\odot}_{\rm infall} > 100\ \mathrm{km\ s^{-1}}$.}
\end{figure}

In panels (a) through (b) of Figure~\ref{figApp:compRem} we show the time evolution of each removal case for both H and H+He removal, for a $M_{\rm ZAMS}=25\Mo$ star. 
Panel (a) depicts the mass evolution, which shows that the default removal scheme of \mesa\ retains more mass than our prescription for both H and H+He removal - consistent with the simulated stars reaching our radius criteria before the (\mesa) composition criteria, and as a result, our core masses are slightly smaller. 
Panel (b) shows the radius evolution of each case. Upon completion of removal, the radius of the H removal cases and the H+He removal cases by our removal scheme is similar to that of their counterparts in the \mesa\ default case (no systematic difference).

In panel (c), we present an HR diagram for the five models. The four removal cases differ greatly in their late stage of evolution, as seen in the separation of the effective temperature $T_{\rm eff}$. This is consistent with the differences in stellar radius shown in panel (b).
For the \mesa\ default removal scheme, the main difference between the two removal cases, H and H+He, is that the H removal case has a higher luminosity than the H+He removal case. The stripped envelope part of the diagram (left) connects to the same main track (right) and then follows a qualitatively similar trajectory in the HR diagram, as in the \mesa \ prescription, both envelope removals are executed consecutively.
For our removal scheme, the two removal cases follow qualitatively different trajectories in the HR diagram. Interestingly, our H+He removal case has a similar trajectory as the \mesa \ default removal scheme, but at a lower luminosity than both cases of the \mesa \ removal scheme. The H removal case in our removal scheme reaches a high luminosity at a lower $\rm T_{eff}$ than all other cases.

Figure~\ref{figApp:compRem} shows several stellar properties of the star when $v^{m < 4 M_\odot}_{\rm{infall}}=100 \kms$. In general, the \mesa \ default removal schemes yield a higher mass for each removal case, H and H+He. By this, the \mesa \ default removal scheme also yields higher binding energy $E_{\rm bin}$, and Fe core mass. 
The temperature and density profiles appear qualitatively similar for the two removal schemes, differing only because of the mass difference. Our Si core is smaller than \mesa \ default in the H removal case, but is roughly the same as \mesa \ default in the H+He removal case. 

Regarding the convective velocity. We find that in both removal cases, the \mesa \ default has a higher convective velocity in the ONe layer, and our removal scheme has a higher convective velocity in the Si layer. In both cases, the convection is sufficient to facilitate the JJEM.

\newpage
%FFFFFFFFFFFFFFFFFFFFFFFFFFFFFFFFFFFFFFFFFFFFFFFFFFFFFFFFFFFFFFFF

%FFFFFFFFFFFFFFFFFFFFFFFFFFFFFFFFFFFFFFFFFFFFFFFFFFFFFFFFFFFFFFFFFFFFFFFFFFFFFFFFFFFFFFFFFFFFFFFFFFFFFFFFF
\clearpage

\setcounter{figure}{0}      
% =========================================
\section{Appendix B: Detailed Numerical Scheme and Consistency}
\label{appendix:numConsistency}
% =========================================
\renewcommand\thefigure{B\arabic{figure}} % more than one figure -> add numbering (\arabic{figure})

%==============
\subsection{On the numerical scheme}
\label{subappendix:NumericalScheme}
% =============

In addition to the default test suite, we used the Ledoux criterion (\citealt{Henyey_etal_1965}) with mixing length theory (MLT) coefficient $\alpha_{\rm MLT}=1.5$ and semi-convection coefficient $\alpha_{\rm SC} = 0.01$. The MLT option is selected to be TDC (time-dependent convection), which converges to the MLT described in \cite{Cox&Giuli_1968} in the long term \citep{Jermyn2023}. For a thorough discussion of the numerical implementation, stability of TDC MLT and comparison to other MLT formulations, see \cite{Jermyn2023}.
We set $\rm Pextra\_factor = 4$ to help stabilize the atmosphere during envelope removal. The \textsc{mesa} parameter `relax\_to\_this\_tau\_factor' is set to $1.5\times 10^6$ to avoid some numerical difficulties when stripping the envelope and is applied to all models, including those with a full envelope. Due to these modifications, the stellar model envelopes are not accurately described. 
This might influence the composition and properties of the outer core. For instance, hydrogen shell burning during the helium burning phase is ignored (e.g., \citealt{IbenRenzini1984, Paxton2011, JonesGore2015}). These are beyond the scope and interest of this paper, as we focus on the inner core layers.

As we manually remove the envelope, we use the default wind scheme and settings. Mass loss by wind is turned on from ZAMS up until the first envelope removal episode (hydrogen envelope removal) with a `Dutch\_scaling\_factor' of $1.0$.

We use the \texttt{mesa\_80} nuclear network, which includes many isotopes of intermediate elements (Oxygen to Silicon) not found in the default \texttt{approx21} massive stars network. \texttt{mesa\_80} was found to give a composition comparable to those from larger networks \citep{Grichener_etal_2025_NuclearNN}. Using larger networks might yield slightly more accurate composition and Y$_{\rm e}$ fractions at the onset of core collapse (e.g., \citealt{Renzo_etal_2024_NucNet}), but it remains computationally demanding. Recent works have presented a possible neural network approach \citep{Grichener_etal_2025_NuclearNN}.

We remove the H envelope when the radius of the photosphere reaches $R_* = 800 R_\odot$, which occurs roughly at the time of central H depletion (He core formation). This envelope-removal scheme qualitatively mimics a binary-interaction case, in which the rapidly expanding photosphere is captured by the companion. The model is then computed until central He depletion, at which point He envelope stripping happens for stellar models with both H and He envelopes stripped. We perform a comparison between our envelope removal scheme and the default scheme used for massive stars in the `20M\_pre\_ms\_to\_core\_collapse' \mesa \ example in Appendix \ref{appendix:removalTiming}. 
 
The envelope removal is done using `relax\_initial\_mass\_to\_remove\_H\_env' and a similar handle `relax\_initial\_mass\_to\_remove\_He\_env', the latter of which we implemented directly into the \mesa \ code and is available in the Zenodo upload. These handles remove mass outside 
\begin{equation}
    M_{\rm remain} = M_{\rm core} + M_{\rm ret}
\end{equation}
where $M_{\rm core}$ is the He core mass or the CO core mass determined by \mesa, where the He core mass is used to remove the H envelope and the CO core mass is used to remove the He envelope. The retained mass $M_{\rm ret} = 5\times 10^{-3} M_\odot$ is the same as the default in \mesa \ and is not altered for the main part of this paper. A more detailed study on the effect of this retained mass will follow in this section. 

We terminate the simulation when the infall velocity at the iron core exceeds $v^{\rm Fe}_{\rm{infall}}=300 \kms$, but choose an earlier time for our analysis of the onset of collapse, $v^{m < 4 M_\odot}_{\rm{infall}}=100 \kms$, i.e., the first time that a point in $m<4 M_\odot$ reaches an infall velocity of $100 \km \s^{-1}$.

A note on the time between simulation termination and the time defined for the onset of collapse. \mesa \ solves temporal advancement of the star under the assumption of hydrostatic equilibrium with hydrodynamic terms \citep{Paxton2011}. Once collapse begins, the iron core quickly compresses in a positive feedback loop. When this is simulated numerically, the escalation between $v^{\rm Fe}_{\rm infall}=100\kms$ and $v^{\rm Fe}_{\rm infall}=300 \kms$ usually corresponds to a single simulation timestep with the default timestep controls. At the onset of collapse, the infall velocity becomes comparable to the convective velocity. Nearing the onset of collapse, we refine the time steps further (by adjusting `time\_delta\_coeff') to avoid the sudden emergence of high infall velocity in the core that terminates the simulation and to ensure several stellar profiles at the onset of collapse.

In any case, convective motion set at the onset of collapse in the Si burning layer would be expected to retain for a time of $t_{conv}\sim\lambda/v_{conv}\approx\frac{300\km}{100\kms}=3\s$ where $\lambda$ is the pressure scale height - well above the free-fall time and the considered explosion timescales.
Additionally, \cite{ShishkinSoker2021} have shown that convective velocities at the Fe/Si interface can further increase beyond the onset of collapse as these layers are compressed and heated. While this will show even larger seed perturbations for the JJEM, we prefer to avoid these fast-infall regimes where the infall velocity is greater than the convective velocity in favor of numerical stability.
We therefore take the convective velocity values that were stable in the seconds leading up to collapse as our typical values and do not explore later simulations times.

\clearpage
%==============
\subsection{Sensitivity to the initial mass}
\label{subappendix:InitialMass}
% =============
%numerical consistency
To test numerical consistency between different simulations, we varied the initial mass of the $15M_\odot$ model by $\pm0.01M_\odot$ and simulated the two resulting models to completion.

\begin{figure*}[h!]
\begin{center}
\includegraphics[trim=0cm 0cm 0cm 0cm,width=1\textwidth]{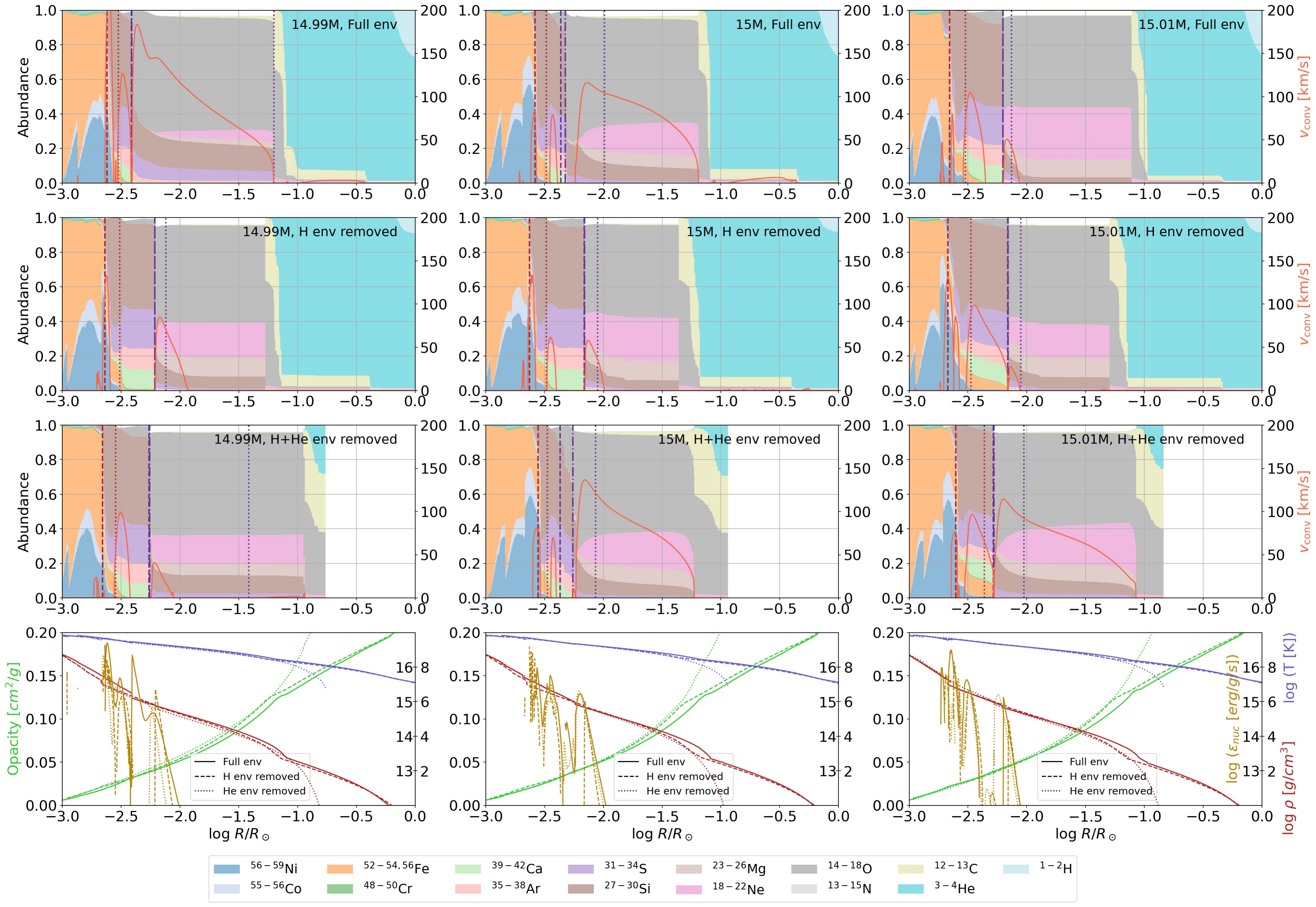} %[trim=left lower right upper]
\caption{Same as Fig.~\ref{fig:comp} but for $M_{\rm ZAMS}=14.99 \Mo$ (left column), $15.00 \Mo$ (centre column), and $15.01\Mo$.}
\label{figApp:comp_15pm}
\end{center}
\end{figure*}

In Figure~\ref{figApp:comp_15pm}, we show the composition and other stellar parameters as a function of radius (the same as in Figure~\ref{fig:comp}) for the three models.
% FFFFFFFFFFFFFFFFFFFFFFFFFFFFFFFFFFFFFFFFFFFFFFFFFFFFFFFFFFFFF
% FFFFFFFFFFFFFFFFFFFFFFFFFFFFFFFFFFFFFFFFFFFFFFFFFFFFFFFFFFFFF

While core masses, total mass fraction, and density/temperature profiles appear qualitatively similar, there are variations in the intermediate layers: 
The $14.99M_\odot$ model has a wide region with $X_{\rm Si,S}>0.1$ and a very gradual transition to the Si/S layer, which is absent in the $15M_\odot$ and $15.01M_\odot$ models.
The $15.01M_\odot$ model also has a wider Si/S layer, similar to that in the H envelope-removed models.
The $15M_\odot /{\rm He}$ model (H+He removed) also has a smaller radius than the $14.99M_\odot /{\rm He}$ and $15.01M_\odot /{\rm He}$ models. 
  
\begin{figure*}[h]
\begin{center}
\includegraphics[trim=0cm 0cm 0cm 0cm,width=1\textwidth]{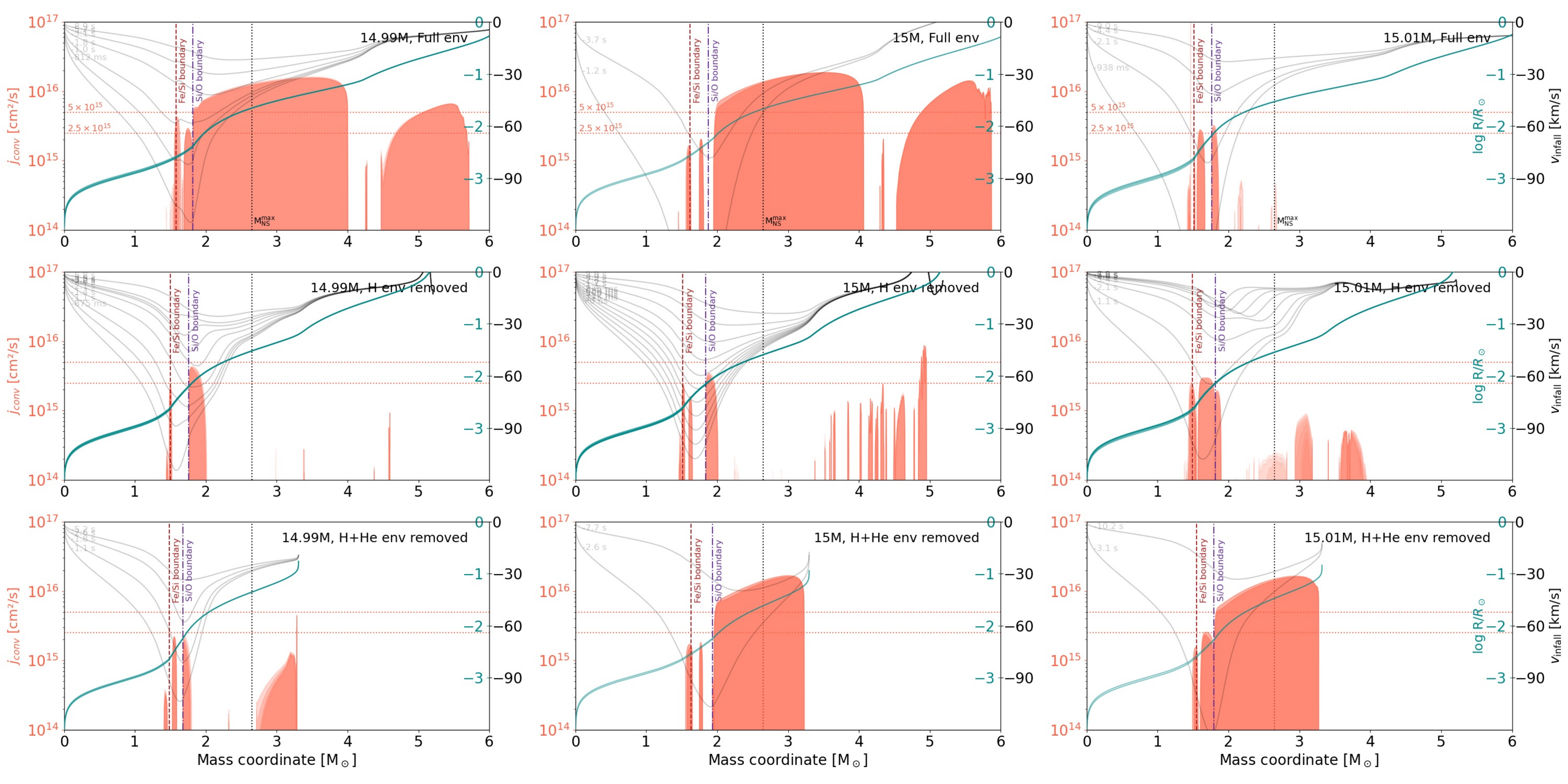} %[trim=left lower right upper]
\caption{Same as Fig.~\ref{fig:ang1} but for $14.99 \Mo$ (left column), $15.00 \Mo$ (centre column), and $15.01\Mo$.}
\label{figApp:ang_15pm}
\end{center}
\end{figure*}

In  Figure~\ref{figApp:ang_15pm}, we present the angular momentum parameter $j(m)$ in the snapshots taken starting from 10 seconds before collapse (same as Figure~\ref{fig:ang1}) for these three models. For the three masses, the hydrogen-removal models exhibit very sporadic and localized convective zones in the oxygen layer. There's a consistent convective zone at the edge of the Si/O boundary as well as one at the Fe/Si interface. The interface convective zones appear in both the full envelope and H+He-removed cases. The $14.99M_\odot$ and $15M_\odot$ models have a large convective zone in the oxygen layer, while the $15.01M_\odot$ model does not (upper row).
The $14.99M_\odot/{\rm He}$ model similarly differs from the $15M_\odot /{\rm He}$ and $15.01M_\odot /{\rm He}$ models, in having a slower, more localized convective zone around $m=(3\pm0.2)M_\odot$, where the oxygen convective zone extends at least $m=2-3M_\odot$.
% FFFFFFFFFFFFFFFFFFFFFFFFFFFFFFFFFFFFFFFFFFFFFFFFFFFFFFFFFFFFF
% FFFFFFFFFFFFFFFFFFFFFFFFFFFFFFFFFFFFFFFFFFFFFFFFFFFFFFFFFFFFF

Differences in the core masses, intermediate layer locations, and mass fractions are expected even for small changes to the initial mass (see Section~\ref{sec:results_models}). These are evidently present in our models, where for a $\pm0.01M_\odot$ difference in the initial mass we get a similar total mass, but slightly different compositions (Figure~\ref{figApp:comp_15pm}) and potentially very different convective zone locations, extent, and velocities approaching collapse (Figure~\ref{figApp:ang_15pm}). This argument extends to the retained mass while removing envelope (See Appendix C on Zenodo).
The conclusion, nonetheless, is that the collapsing core has the required convective zones to seed the instabilities that feed the intermittent stochastic accretion disk that launches the jittering jets.  

\clearpage

\end{document}